\documentclass[journal,twoside]{IEEEtran}
\usepackage{cite}

\ifCLASSINFOpdf
  \usepackage[pdftex]{graphicx}
  \graphicspath{{fig/}}
  \DeclareGraphicsExtensions{.pdf,.jpeg,.png,.eps}
\else
\fi
\usepackage{amsmath}
\usepackage{array}

\ifCLASSOPTIONcompsoc
  \usepackage[caption=false,font=normalsize,labelfont=sf,textfont=sf]{subfig}
\else
  \usepackage[caption=false,font=footnotesize]{subfig}
\fi
\usepackage{url}

\usepackage[dvipsnames]{xcolor}
\usepackage{xifthen}

\usepackage[binary-units=true, per-mode=symbol]{siunitx}
\usepackage[acronym]{glossaries}
\usepackage{xspace}
\usepackage{tabularx}
\usepackage{booktabs}

\usepackage{multirow}
\usepackage[para]{threeparttable}
\usepackage{tikz}
\usepackage[shortcuts]{extdash}

\DeclareSIUnit\IOP{iop}
\DeclareSIUnit\FLOP{flop}
\DeclareSIUnit\FLOPs{\FLOP\per\second}
\DeclareSIUnit\FLOPsW{\FLOPs\per\watt}
\DeclareSIUnit\GFLOPs{\giga\FLOPs}
\DeclareSIUnit\GFLOPsW{\giga\FLOPsW}
\DeclareSIUnit\TFLOPs{\tera\FLOPs}
\DeclareSIUnit\TFLOPsW{\tera\FLOPsW}

\newcolumntype{C}{>{\centering\arraybackslash}X}
\newcolumntype{R}{>{\raggedleft\arraybackslash}X}
\newcolumntype{x}[1]{>{\centering\arraybackslash\hspace{0pt}}p{#1}}
\newcolumntype{k}[1]{S[
  table-format=#1,
  detect-weight,
  input-symbols={()}
]}

\newcommand{\secref}[1]{{Section\,\ref{#1}}}
\newcommand{\figref}[1]{{Fig.\,\ref{#1}}}
\newcommand{\tabref}[1]{{Table\,\ref{#1}}}

\newcommand{\fpstd}{IEEE~754\xspace}
\newcommand{\fpstdnew}{IEEE~754\=/2008\xspace}
\newcommand{\riscv}{RISC\=/V\xspace}
\newcommand{\gf}{\textsc{Glo\-bal\-found\-ries}~22FDX\xspace}
\newcommand{\wrt}{w.r.t. }
\newcommand{\wfpu}{\texttt{w}_{fpu}}
\newcommand{\dbit}[1]{{#1}\=/\si{\bit}}

\newacronym{tpfpu}{TP-FPU}{transprecision floating-point unit}
\newacronym{fpu}{FPU}{floating-point unit}
\newacronym{fp}{FP}{floating-point}
\newacronym{simd}{SIMD}{single instruction multiple data}
\newacronym{hpc}{HPC}{High Performance Computing}
\newacronym{iot}{IoT}{Internet of Things}
\newacronym{fma}{FMA}{fused multiply-add}
\newacronym{nan}{NaN}{not a number}
\newacronym{isa}{ISA}{instruction set architecture}
\newacronym{fpga}{FPGA}{field-programmable gate array}
\newacronym{asic}{ASIC}{application-specific integrated circuit}
\newacronym{unum}{UNUM}{universal number}
\newacronym{lns}{LNS}{logarithmic nubmer system}
\newacronym{fbb}{FBB}{forward body bias}
\newacronym{mac}{MAC}{multiply-accumulate}

\DeclareSIUnit \gateequivalent {GE}

\begin{document}

%
\title{FPnew: An Open-Source Multi-Format Floating-Point Unit Architecture for Energy-Proportional Transprecision Computing}
%
%
%

\author{Stefan~Mach,
        Fabian~Schuiki,
        Florian~Zaruba,~\IEEEmembership{Student~Member,~IEEE,}
        and~Luca~Benini,~\IEEEmembership{Fellow,~IEEE}
\thanks{S. Mach, F. Schuiki, and F. Zaruba are with the Integrated Systems Laboratory, ETH Zürich, 8092 Zürich, Switzerland (e-mail: \{smach, fschuiki, zarubaf\}@iis.ee.ethz.ch).}
\thanks{L. Benini  is with the  Integrated  Systems  Laboratory,  ETH Zürich, 8092 Zürich, Switzerland, and also with the Department of Electrical, Electronic, and Information Engineering Guglielmo Marconi, University of Bologna, 40136 Bologna, Italy (e-mail: lbenini@iis.ee.ethz.ch).}
\thanks{This work has received funding from the European Union’s Horizon 2020 research and innovation program under grant agreement No 732631, project ``OPRECOMP''.
Supported in part by the European Union's H2020 EPI project under grant agreement number 826647.
}}%

\maketitle

\begin{abstract}
  The slowdown of Moore's law and the power wall necessitates a shift towards finely tunable precision (a.k.a. transprecision) computing to reduce energy footprint.
  Hence, we need circuits capable of performing floating-point operations on a wide range of precisions with high energy-proportionality.
  We present FPnew, a highly configurable open-source transprecision floating-point unit (TP-FPU) capable of supporting a wide range of standard and custom FP formats.
  To demonstrate the flexibility and efficiency of FPnew in general-purpose processor architectures, we extend the RISC-V ISA with operations on half-precision, bfloat16, and an 8bit FP format, as well as SIMD vectors and multi-format operations.
  Integrated into a 32-bit RISC-V core, our TP-FPU can speed up execution of mixed-precision applications by 1.67x w.r.t. an FP32 baseline, while maintaining end-to-end precision and reducing system energy by 37\%.
  We also integrate FPnew into a 64-bit RISC-V core, supporting five FP formats on scalars or 2, 4, or 8-way SIMD vectors.
  For this core, we measured the silicon manufactured in Globalfoundries 22FDX technology across a wide voltage range from 0.45V to 1.2V.
  The unit achieves leading-edge measured energy efficiencies between 178 Gflop/sW (on FP64) and 2.95 Tflop/sW (on 8-bit mini-floats), and a performance between 3.2 Gflop/s and 25.3 Gflop/s.
\end{abstract}

\begin{IEEEkeywords}
Floating-Point Unit, RISC-V, Transprecision Computing, Multi-Format, Energy-Efficient.
\end{IEEEkeywords}

%
\IEEEpeerreviewmaketitle


\section{Introduction}
\label{sec:intro}

\IEEEPARstart{T}{he} last decade has seen explosive growth in the quest for energy-efficient architectures and systems.
An era of exponentially improving computing efficiency-driven mostly by CMOS technology scaling is coming to an end as Moore's law falters.
The obstacle of the so-called thermal- or power-wall is fueling a push towards computing paradigms, which hold energy efficiency as the ultimate figure of merit for any hardware design.

At the same time, rapidly evolving workloads such as Machine Learning is the focus of the computing industry and always demand higher compute performance at constant or decreasing power budgets, ranging from the data-center and \gls{hpc} scale down to the \gls{iot} domain.
In this environment, achieving high energy efficiency in numerical computations requires architectures and circuits which are fine-tunable in terms of precision and performance.
Such circuits can minimize the energy cost per operation by adapting both performance and precision to the application requirements in an agile way.
The paradigm of ``transprecision computing'' \cite{malossi2018transprecision} aims at creating a holistic framework ranging from algorithms and software down to hardware and circuits which offer many knobs to fine-tune workloads.

The most flexible and dynamic way of performing numerical computations on modern systems is \gls{fp} arithmetic.
Standardized in \fpstd, it has become truly ubiquitous in most computing domains: from general-purpose processors, accelerators for graphics computations (GPUs) to \gls{hpc} supercomputers, but also increasingly in high-performance embedded systems and ultra-low-power microcontrollers.
While fixed-point computation, which usually makes use of integer datapaths, sometimes offers an efficient alternative to \gls{fp}, it is not nearly as flexible and universal.
Domain-specific knowledge by human experts is usually required to transform \gls{fp} workloads into fixed-point, as numerical range and precision trade-offs must be managed and tracked manually.
\fpstd's built-in rounding modes, graceful underflow, and representations for infinity are there to make \gls{fp} arithmetic more robust and tolerant to numerical errors \cite{kahan1998java}.
Furthermore, many applications such as scientific computing with physical and chemical simulations are infeasible in fixed-point and require the dynamic range, which \gls{fp} offers.

\gls{fp} precision modulation as required for efficient transprecision computing has been limited to the common ``double'' and ``float'' formats in CPUs and GPUs in the past.
However, a veritable ``Cambrian Explosion'' of \gls{fp} formats, e.g. Intel Nervana's Flexpoint \cite{koster2017flexpoint}, Microsoft Brainwave's \dbit{9} floats \cite{chung2018serving}, the Google TPU's \dbit{16} ``bfloats'' \cite{jouppi2017datacenter}, or \textsc{Nvidia}'s \dbit{19} TF32
, implemented in dedicated accelerators such as Tensor Cores \cite{amit2018extreme}, shows that new architectures with extreme transprecision flexibility are needed for \gls{fp} computation, strongly driven by machine learning algorithms and applications.
Our goal is to create a flexible and customizable \gls{tpfpu} architecture that can be utilized across a wide variety of computing systems and applications.

In order to leverage such transprecision-enabled hardware, there must, of course, also be support and awareness across the entire software stack.
An \gls{isa} forms the interface between hardware and software.
\riscv \cite{riscvspec} is an open-source \gls{isa} which natively supports computation on the common ``double'' and ``float'' formats.
Furthermore, the \gls{isa} explicitly allows non-standard extensions where architects are free to add instructions of their own.
Lately, \riscv has gained traction in both industry and academia due to its open and extensible nature with growing support from hardware and software projects.
In this work, we leverage the openness and extensibility of the \riscv \gls{isa} by adding extensions for operations on additional \gls{fp} formats not found in current \riscv processor implementations \cite{tagliavini2019extension}.


In this work, we also demonstrate a fully functional silicon implementation of a complete open-source \gls{tpfpu} inside a \riscv application-class core in a \SI{22}{\nm} process \cite{mach20190}.
The taped-out architecture supports a wide range of data formats including \fpstd{} double (FP64), single (FP32), and half-precision floats (FP16), as well as \dbit{16} bfloats (FP16alt) and a custom \dbit{8} format (FP8), initially introduced in \cite{tagliavini2018transprecision}.
Furthermore, there is full support for \gls{simd} vectorization, as well as vectorial conversions and data packing.

To summarize, our contributions are:
\begin{enumerate}
  \item The design of a highly configurable architecture for a \acrlong{tpfpu} written in SystemVerilog.
        All standard \riscv operations are supported along with various additions such as \gls{simd} vectors, multi-format \gls{fma} operations or convert-and-pack functionality to dynamically create packed vectors.
        The unit is fully open-source and thus extensible to support even more functions.
  \item Extensions to the \riscv \gls{isa} to support transprecision \gls{fp} operations on FP64, FP32, FP16, FP16alt, and FP8 \cite{tagliavini2018transprecision}.
        Programmers can leverage transprecision through the use of standard operations in high-level programming languages and make use of compiler-enabled auto-vectorization, or make further optimization using compiler-intrinsic function calls to transprecision instructions \cite{tagliavini2019extension}.
  \item Integration of the \gls{tpfpu} into RI5CY\cite{gautschi2017near}, a \dbit{32} embedded \riscv processor core.
        An application case study shows that using our transprecision \gls{isa} extension can achieve a $1.67\times$ speedup to an FP32 baseline without sacrificing any precision in the result.
        Furthermore, the processor energy required to complete the workload is reduced by $37\%$.
  \item Integration of the \gls{tpfpu} into Ariane\cite{zaruba2019ariane}, a \dbit{64} application-class \riscv processor core and subsequent silicon implementation in \gf \cite{mach20190}.
        Energy and performance measurements of the manufactured silicon confirm the substantial energy proportionality and leading-edge energy efficiency of our architecture.
        We perform a detailed breakdown of per-instruction energy cost, the gains of vectorization, and an evaluation of the voltage/frequency scaling and body biasing impact on the manufactured silicon.
        Our design surpasses the SOA of published \gls{fpu} designs in both flexibility and efficiency.
\end{enumerate}

The rest of the paper is organized as follows:
\secref{sec:arch} describes in-depth the requirements and architecture of the proposed \gls{tpfpu}.
\secref{sec:impl} outlines our work on transprecision \gls{isa} extensions, the implementation of the hardware unit into the two processor cores, and the full implementation into silicon.
\secref{sec:results} contains a transprecision case study performed on the RI5CY core system as well as the silicon measurement results of the Ariane core system.
The last sections of this paper contrast our work with related works and provide a summary of our work.

\section{Architecture}
\label{sec:arch}

FPnew is a flexible, open-source hardware IP block that adheres to \fpstd{} standard principles, written in SystemVerilog.
The aim is to provide \gls{fp} capability to a wide range of possible systems, such as general-purpose processor cores as well as domain-specific accelerators.

\subsection{Requirements}

To address the needs of many possible target systems, applications, and technologies, FPnew had configurability as one of the driving factors during its development.
The ease of integration with existing designs and the possibility of leveraging target-specific tool flows was also a guiding principle for the design.
We present some key requirements that we considered during the design of the unit:

\subsubsection{FP Format Encoding}

\begin{figure}
    \centering
    \includegraphics[width=0.7\linewidth]{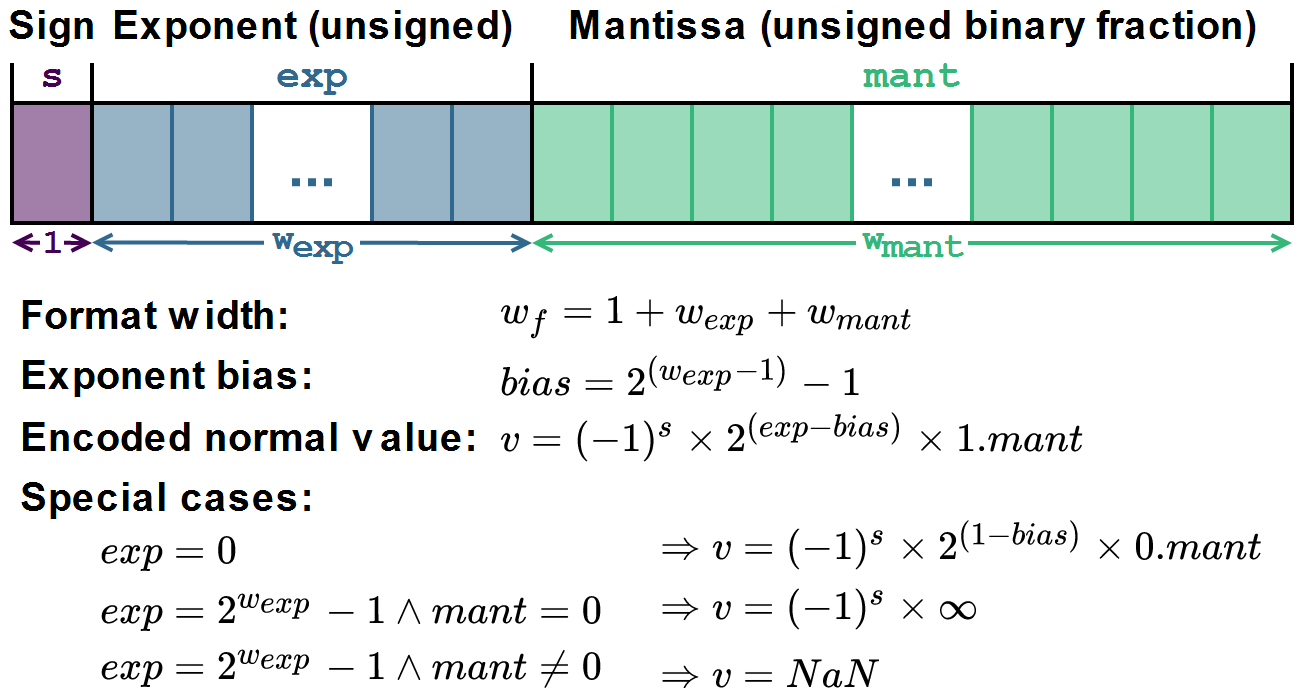}
    \vspace{-0.2cm}
    \caption{\gls{fp} format encoding as specified by \fpstd{} and its interpretation.}
    \vspace{-0.3cm}
    \label{fig:fpfmt}
\end{figure}

As outlined in \secref{sec:intro}, it is becoming increasingly attractive to add custom \gls{fp} formats (often narrower than \SI{32}{\bit}) into a wide range of systems.
While many of the systems mentioned earlier abandon standard compliance for such custom formats in pursuit of optimizations in performance or circuit complexity, general-purpose processors are generally bound to adhere to the \fpstd{} standard. 
As such, the \gls{tpfpu} is designed to support any number of arbitrary \gls{fp} formats (in terms of bit width) that all follow the principles for \fpstdnew{} \emph{binary} formats, as shown in \figref{fig:fpfmt}.

\subsubsection{Operations}

To provide a complete FPU solution, we aim at providing the general operations mandated by \fpstd{}, supporting arithmetic operations, comparisons, and conversions.
Most notably, the \gls{fma} operation that was first included in a processor in 1990 \cite{montoye1990design} has since been added to \fpstdnew{} and is nowadays ubiquitous in efficient AI and BLAS-type kernels.
It computes $(a \times b) + c$ with only one final rounding step.
We aim at natively supporting at least all \gls{fp} operations specified in the \riscv{} ISA.

Furthermore, for implementations supporting more than one \gls{fp} format, conversions among all supported \gls{fp} formats and integers are required.
Non-standard multi-format arithmetic is also becoming more common, such as performing the multiplication and accumulation in an \gls{fma} using two different formats in tensor accelerators \cite{jouppi2017datacenter, amit2018extreme}.

\subsubsection{SIMD Vectors}

Nowadays, most general-purpose computing platforms offer \gls{simd} accelerator extensions, which pack several narrow operands into a wide datapath to increase throughput.
While it is possible to construct such a vectorized wide datapath by duplicating entire narrow \glspl{fpu} into vector lanes, operations would be limited to using the same narrow width.
Flexible conversions amongst \gls{fp} types are crucial for efficient on-the-fly precision adjustment in transprecision applications \cite{mach2018transprecision} and require support for vectored data.
The architecture of the \gls{tpfpu} thus must be able to support this kind of vectorization to support multi-format operations on \gls{simd} vectors.

\subsubsection{Variable Pipeline Depths}


In order to be performant and operate at high speeds, commonly used operations inside an \gls{fpu} require pipelining.
However, pipeline latency requirements for \gls{fp} operations are very dependent on the system architecture and the choice of implementation technology.
While a GPU, for example, will favor a minimum area implementation and is capable of hiding large latencies well through its architecture, the impact of operation latency can be far more noticeable in an embedded general-purpose processor core \cite{mach2018transprecision}.

As such, the \gls{tpfpu} must not rely on hard-coding any specific pipeline depths to support the broadest possible range of application scenarios.
As circuit complexity differs significantly depending on the operation and \gls{fp} format implemented, the number of registers shall be configurable independently for each.

\subsubsection{Design Tool Flow}


Target-specific synthesis flows (e.g. for \gls{asic} or \gls{fpga} technologies) differ in available optimized blocks, favoring inferrable operators over direct instantiation.
Synthesis tools will pick optimal implementations for arithmetic primitives such as DSP slices in \glspl{fpga} or Wallace-Tree based multipliers for \glspl{asic} with high timing pressure.
As available optimizations also differ between targets, the unit is described in a way to enable automatic optimizations, including clock-gating and pipelining, wherever possible.

\subsection{Building Blocks}

\begin{figure*}
    \centering
    \includegraphics[width=\linewidth]{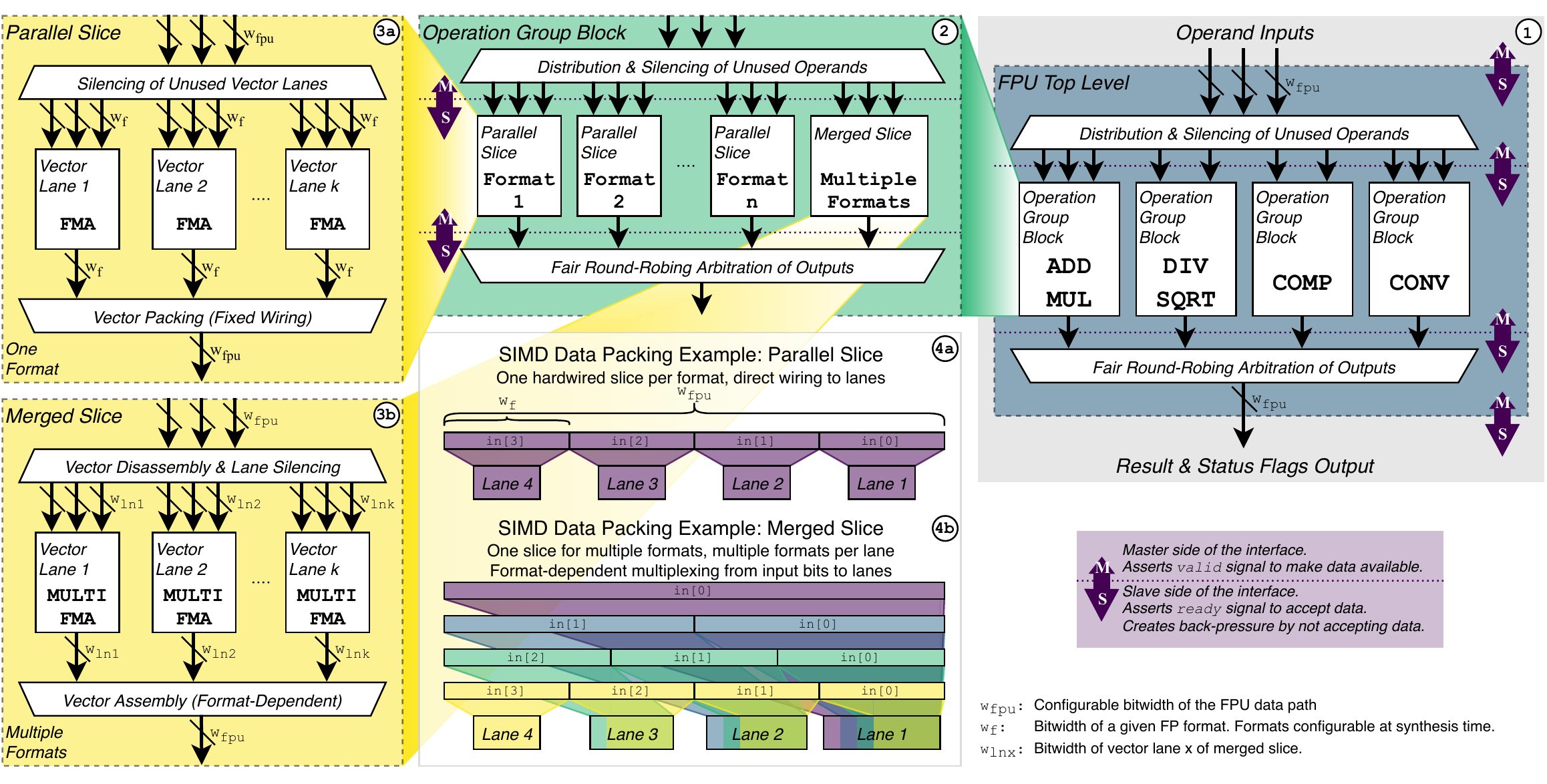}
    \vspace{-0.8cm}
    \caption{Datapath block diagram of the \gls{tpfpu} with its underlying levels of hierarchy.
             It supports multiple \gls{fp} formats, and the datapath width is configurable.
            }
    \vspace{-0.5cm}
    \label{fig:fpu_top}
\end{figure*}

In the following we present a general architectural description of our \gls{tpfpu}, shown in \figref{fig:fpu_top}.
Concrete configurations chosen for the integration into processor cores and the implementation in silicon are discussed in Sections~\ref{sec:impl}~and~\ref{sec:results}.

\subsubsection{FPU Top Level}

At the top level of the \gls{tpfpu} (see \figref{fig:fpu_top}\=/1), up to three \gls{fp} operands can enter the unit per clock cycle, along with control signals that determine the type of operation as well as the format(s) involved.
One \gls{fp} result leaves the unit along with the status flags raised by the current operation according to \fpstdnew{}.
The width of the input and output operands is parametric and will be henceforth referred to as the \emph{unit width} ($\wfpu$).

Input operands are routed towards one of four operation group blocks, each dedicated to a class of instructions.
Arbiters feed the operation group outputs towards the output of the unit.
As only one operation group block can receive new data in any given clock cycle, clock and datapath gating can be employed to silence unused branches of the \gls{fpu}, thereby eliminating spurious switching activity.

\subsubsection{Operation Group Blocks}


The four operation group blocks making up the \gls{tpfpu} are as follows:
\begin{itemize}
    \item \emph{ADDMUL}: addition, multiplication and \gls{fma}
    \item \emph{DIVSQRT}: division and square root
    \item \emph{COMP}: comparisons and bit manipulations
    \item \emph{CONV}: conversions among \gls{fp} formats as well as to and from integers
\end{itemize}

Each of these blocks forms an independent datapath for operations to flow through (see \figref{fig:fpu_top}\=/2).
When multiple \gls{fp} formats are present in the unit, the blocks can host several slices that are either implemented as format-specific (parallel) or multi-format (merged).
In the parallel case, each slice hosts a single \gls{fp} format, giving the broadest flexibility in terms of path delay and latency, as each slice can contain its internal pipeline.
Inactive format slices can be clock-gated and silenced.
Note, however, that this flexibility comes at an increased area cost since the datapath is unique for each format.
In contrast, a merged slice can lower total area costs by sharing hardware, using a single slice that houses multiple formats at reduced flexibility.
Furthermore, merging may incur energy and latency overheads due to small formats reusing the same over-dimensioned datapath, with the same pipeline depth for all formats.

\subsubsection{Format Slices}

Format slices host the functional units which perform the operations that the block is specialized in.

A \gls{simd} vector datapath can be created if the format can be packed into the unit width ($\wfpu \geq 2\times\texttt{w}_{f}$).
In this case slices will host multiple vector lanes, denoted $\texttt{lane[1]} \dots \texttt{lane[k]}$.

In the parallel case (see \figref{fig:fpu_top}\=/3a), the lanes are duplicate instances of the same functional unit, and the number and width of lanes is determined as follows.
\begin{align*}
    \texttt{k}_{parallel} &= \left\lfloor\frac{\wfpu}{\texttt{w}_{f}}\right\rfloor\\
    \texttt{w}_{lane, parallel} &= \texttt{w}_{f}
\end{align*}

In the merged case (see \figref{fig:fpu_top}\=/3b), the total number of lanes is determined by the smallest supported format, and the width of each lane depends on the containing formats.
Individual lanes within merged slices have the peculiar property of differing in bit width, and each lane needs support for a different set of formats (see \figref{fig:fpu_top}\=/4b).
\begin{align*}
    \texttt{k}_{merged} &= \left\lfloor\frac{\wfpu}{\min\limits_{\forall format \in slice}\texttt{w}_{f}}\right\rfloor\\
    \texttt{w}_{lane[i], merged} &= \max\limits_{\forall format \in slice}\texttt{w}_{f}|_{\texttt{w}_{f} \leq \frac{\wfpu}{i}}
\end{align*}

Depending on whether the current operation is scalar or vectored, either one or several lanes are used to compute the slice's result while unused lanes are silenced.
The merged slices in the CONV block require a more complex data distribution and collection scheme for \gls{simd} vectors as input and output format widths can differ.
Furthermore, it is possible to cast two scalar \gls{fp} operands and insert them as elements of vectors for the dynamic creation of vectors at runtime.
If \gls{simd} is disabled, there is only one lane per slice.

\subsubsection{Functional Units}

The functional units within a slice can either be fully pipelined or use a blocking (e.g., iterative) implementation.

The ADDMUL block uses fully pipelined \gls{fma} units compliant with \fpstdnew{}, implemented using a single-path architecture \cite{montoye1990design, MullerEtAl2018}, providing results within $1/2$ ulp.
A multi-format version of the \gls{fma} unit is used in merged slices, supporting mixed-precision operations that use multiple formats.
Namely, the multiplication is done in the source format \texttt{src\_fmt} while the addition is done using the destination format \texttt{dst\_fmt}, matching the C-syle function prototype $\texttt{\emph{dst\_fmt} fma(\emph{src\_fmt}, \emph{src\_fmt}, \emph{dst\_fmt)}}$.

In the DIVSQRT block, divisions and square roots are computed using an iterative non-restoring divider.
The iterative portion of the unit computes three mantissa bits of the result value per clock cycle and is implemented as a merged slice.
The number of iterations performed can be overridden to be fewer than needed for the correctly rounded result to trade throughput for accuracy in transprecision computing.
The operational unit in the COMP block consists of a comparator with additional selection and bit manipulation logic to perform comparisons, sign manipulation as well as \gls{fp} classification.
Lastly, the CONV block features multi-format casting units that can convert between any two FP formats, from integer to FP formats, and from FP to integer formats.

\subsubsection{Unit Control Flow}

While only one operation may enter the \gls{fpu} per cycle, multiple values coming from paths with different latencies may arrive at the slice outputs in the same clock cycle.
Resulting data from all slices and blocks are merged using fair round-robin arbitration.
In order to stall internal pipelines, a simple synchronous \emph{valid-ready} handshaking protocol is used within the internal hierarchies as well as on its outside interface of the unit.

As the unit makes heavy use of handshaking, data can traverse the \gls{fpu} without the need for apriori knowledge of operation latencies.
Fine-grained clock gating based on handshake signals can thus occur within individual pipeline stages, silencing unused parts and ``popping'' pipeline bubbles by allowing data to catch up to the stalled head of a pipeline.
Coarse-grained clock gating can be used to disable operation groups or the entire \gls{tpfpu} if no valid data is present in the pipeline.

\subsubsection{Configuration}

The \gls{tpfpu} can be configured in many ways using SystemVerilog parameters and packages alone.
Particularly, the most central configuration options are the following:
    (i)
        Custom formats can be defined containing any number of exponent and mantissa bits and are not limited to power-of-two format widths as is customary in traditional computing systems%
        \footnote{In order to meaningfully interpret bit patterns as \gls{fp} values according to \fpstd{}, a format should contain at least \SI{2}{\bit} each of exponent and mantissa. The SystemVerilog language does not guarantee support for signal widths above \SI[parse-numbers=false]{2^{16}}{\bit}, which is far beyond the reasonable use case of a \gls{fp} format.}.
        Formats are treated according to \fpstdnew{} (see \figref{fig:fpfmt}) and support all standard rounding modes.
        Virtually any number of formats can be supported within the unit.
    (ii)
        Separately for every operation group, each format can be implemented as either a parallel or a merged slice.
        The generation of hardware for any given format in the operation group can also be disabled completely.
    (iii)
        \gls{simd} vectors can be enabled globally for all formats with $\texttt{w}_{f}\leq\wfpu/2$.
        Notably, $\wfpu$ can be chosen much wider than the largest supported format, creating a \gls{simd} \gls{fpu} that can be used in vector accelerators, for example.
    (iv)
        The number of pipeline stages can be freely set for each format and operation group, with merged slices using the highest number of stages of any containing format.
        As pipeline registers are inserted at predefined locations of the functional units, retiming features of synthesis tools might be required to optimize these registers' placement.






\section{Integrating FPnew in RISC-V Cores}
\label{sec:impl}

The \gls{tpfpu} has been integrated into several designs, and this work focusses on the implementation within \riscv{} processor cores.
In order to leverage transprecision computing on \riscv{} platforms, we have extended the \gls{isa} with special instructions.
We integrated the unit into RI5CY, a \dbit{32} low-power core, and Ariane, a \dbit{64} application-class processor.

\subsection{ISA Extensions}

The \riscv{} \gls{isa} offers ample opportunities for extensions with custom operations.
Therefore, we add non-standard \gls{fp} formats and instructions to enable transprecision computing and fully leverage our \gls{tpfpu} in general-purpose processors.

\subsubsection{FP Formats}

In addition to the \fpstd{} \emph{binary32} and \emph{binary64} formats included in \riscv{} `F' and `D' standard extensions, respectively, we also offer smaller-than-\SI{32}{\bit} formats proposed in \cite{tagliavini2018transprecision}.
The available \gls{fp} formats in our implementations are:
\begin{itemize}
    \item \textbf{binary64} (FP64): \fpstd double-precision (11, 52)
    \item \textbf{binary32} (FP32): \fpstd single-precision (8, 23)
    \item \textbf{binary16} (FP16): \fpstd half-precision (5, 10)
    \item \textbf{binary16alt} (FP16alt): custom half-precision (8, 7)\footnote{This format has been popularized under the name \emph{bfloat16}. Our implementation differs from \emph{bfloat16} insofar we always follow \fpstd prinicples regarding denormal and infinity values, \gls{nan}, and support all rounding modes.}
    \item \textbf{binary8} (FP8): custom quarter-precision minifloat (5, 2)
\end{itemize}
Data in all these formats are treated analogously to standard \riscv{} \gls{fp} formats, including the support for denormals, \gls{nan} and the \acrshort{nan}-boxing of narrow values inside wide \gls{fp} registers.

\subsubsection{Operations}

The new operations can be roughly grouped into three parts, namely \emph{scalar}, \emph{vectorial}, and \emph{auxiliary} extensions \cite{tagliavini2019extension}.

\paragraph{Scalar Instructions}
The scalar extensions map all the operations found in, e.g., in the `F' standard extension, such as arithmetic, comparisons, conversions, and data movement to the newly introduced formats.
Conversions among all supported \gls{fp} types were added to enable efficient runtime precision-scaling in transprecision applications.

\paragraph{Vectorial Instructions}
We add \gls{simd} capabilities on all supported \gls{fp} formats that are narrower than the \gls{fp} register file size (\emph{FLEN} in \riscv parlance).
Thus, in a core with support for up to FP32 ($FLEN=32$), a packed vector of 2 $\times$ FP16 is possible.
Our \gls{isa} extension includes vectorial versions of all scalar instructions.
Furthermore, we add \emph{vector-scalar} versions of these operations where the second operand is a scalar.
The scalar operand is replicated to all elements of the input vector, allowing \gls{simd} matrix product computations without the need for transposing one matrix in memory, for example.

Converting between two formats require special care for vectors as their length can differ.
In a system with $FLEN=32$, converting a vector of 2$\times$FP16 to FP8 yields only two elements (\SI{16}{\bit}) of the 4-element (\SI{32}{\bit}) destination.
Conversely, converting a vector of 4 $\times$ FP8 to FP16 would produce a \dbit{64} result, which does not fit the register.
Therefore, we provide separate instructions to use the lower or upper part of a vector for vectorial conversions, allowing for flexible precision scaling in transprecision applications.

\paragraph{Auxiliary Instructions}
Some non-standard operations to address the needs of transprecision computing systems complete our \gls{isa} extensions.
For example, we add an expanding \gls{fma} operation, which performs the sum on a larger \gls{fp} format than the multiplication, mapping to multi-precision operations of the merged \gls{fma} slice of the \gls{tpfpu} architecture.
Our \emph{cast-and-pack} instructions convert two scalar operands using the vectorial conversion hardware and subsequently pack them into elements of the result vector.

\subsubsection{Encoding}
The encoding of these new instructions was implemented as \riscv brown-field non-standard \gls{isa} extensions\footnote{\url{https://iis-git.ee.ethz.ch/smach/smallFloat-spec/blob/v0.5/smallFloat_isa.pdf}}.
As the scalar instructions only introduce new formats, the encoding of the standard \riscv \gls{fp} instructions is reused and adapted.
We use a reserved format encoding to denote the FP16 format and reuse the encodings in the quad-precision standard extension `Q' to denote FP8, as we are not targeting any \riscv processor capable of providing \dbit{128} \gls{fp} operations.
Operations on FP16alt are encoded as FP16 with a reserved rounding mode set in the instruction word.
Vectorial extensions make use of the vast unused space in the integer operation opcode space, similarly to the encoding of DSP extensions realized for the RI5CY core \cite{gautschi2017near}.
Auxiliary instructions are encoded either in unused \gls{fp} or integer operation opcode space, depending on whether they operate on scalars or vectors.

\subsubsection{Compiler Support}
Programmers require high-level support for novel features in computer architectures in order to make efficient use of them.
As such, the formats and operations mentioned above were added into the \riscv GCC compiler toolchain to allow for native support of transprecision operations in user programs \cite{tagliavini2019extension}.
Custom formats can be used like the familiar native \gls{fp} types in the C/C++ programming language, e.g. the new \texttt{float8} C type denotes an FP8 variable.

\subsection{RI5CY with Transprecision FPU}
\label{sec:implriscy}

RI5CY is an open-source \dbit{32}, four stage, in-order \riscv RV32IMFC processor\footnote{https://github.com/pulp-platform/riscv}.
This small core is focussed on embedded and DSP applications, featuring several custom non-standard \riscv extensions for higher performance, code density, and energy efficiency \cite{gautschi2017near}.
With this core, we want to showcase non-standard transprecision operations within a low-power MCU-class open-source \riscv core, which has gained broad industry adoption\footnote{\url{https://www.openhwgroup.org}}.

\subsubsection{ISA Extension Support}
RI5CY supports the \riscv `F' standard \gls{isa} extension, which mandates the inclusion of 32 \dbit{32} \gls{fp} registers.
The core offers the option to omit the \gls{fp} registers and host \gls{fp} data within the general-purpose register file to conserve area and reduce data movement\footnote{At the time of writing, this extension is being considered as an official \riscv extension `Zfinx,' but specification work is not completed.}.

We add support for operations on FP16 and FP16alt, including packed \gls{simd} vectors.
By reusing the general-purpose register file for \gls{fp} values, we can leverage the \gls{simd} shuffling functionality present in the integer datapath through the custom DSP extensions.
Support for both cast-and-pack as well as expanding FMA is added to the core as well.

\subsubsection{Core Modifications}
To handle these new instructions, we extend the processor's decoder with the appropriate instruction encodings.
\riscv{} requires so-called \acrshort{nan}-boxing of narrow FP values where all unused higher order bits of a \gls{fp} register must be set to logic high.
We extend the load/store unit of the core to allow for one-extending scalar narrow \gls{fp} data by modifying the preexisting sign-extension circuitry.
We do not enforce the checking of \acrshort{nan}-boxing in the operation units, however, in order to be able to treat \gls{simd} data as scalars if needed.
Other than replacing RI5CY's FP32 \gls{fpu} with the \gls{tpfpu}, the changes to the core itself are not very substantial compared to the infrastructure already in place for the `F' extension.

\subsubsection{FPU Configuration}

\begin{table}
  \centering
  \begin{threeparttable}
    \centering
    \caption{The configuration of the \gls{tpfpu} as implemented into the RI5CY core. The FPU width is $\wfpu = \SI{32}{\bit}$.}
    \label{tab:config_riscy}
    \begin{tabularx}{\linewidth}{@{} Xllll @{}}
    \toprule
    \textbf{Format} &
    \multicolumn{4}{l}{\textbf{Implementation} (number of cycles, number of lanes)}\\
    &
    ADDMUL &
    DIVSQRT &
    COMP &
    CONV \\
    \midrule
    FP32    & merged (1,1) & disabled (-,-) & parallel (1,1) & merged (1,2)\\
    FP16    & merged (1,2) & disabled (-,-) & parallel (1,2) & merged (1,2)\\
    FP16alt & merged (1,2) & disabled (-,-) & parallel (1,2) & merged (1,2)\\
    \bottomrule
    \end{tabularx}
  \end{threeparttable}
  \vspace{-0.5cm}
\end{table}

We enable support for the above formats without adding any extra pipeline stages, as shown in \tabref{tab:config_riscy}.
Low-power MCUs target relatively relaxed clock targets, such that \gls{fp} operations can complete within a single cycle.
As $XLEN=32$ and \gls{fp} operations use the general purpose register file, $\wfpu$ is set to \SI{32}{\bit}.

The ADDMUL block is implemented as a merged multi-format slice to allow for multi-format operations among FP16[alt] and FP32.
The DIVSQRT block has been disabled as we do not utilize it for our case study and to demonstrate the fine-grained configurability of the \gls{tpfpu}.
The CONV block uses two \dbit{32} lanes in a merged slice to enable cast-and-pack operations from two FP32 operands.

\subsection{Ariane with Transprecision FPU}
\label{sec:implariane}
Ariane is an open-source \dbit{64}, six stage, partially in-order \riscv RV64GC processor\footnote{https://github.com/pulp-platform/ariane}.
It has full hardware support for running an operating system as well as private instruction and data caches.
To speed up sequential code, it features a return address stack, a branch history table, and a branch target buffer \cite{zaruba2019ariane}.
We aim at bringing a full transprecision computing system to silicon with this core, with support for energy-proportional computation supporting many formats.

\subsubsection{ISA Extension Support}
Ariane supports the \riscv `F' and `D' standard \gls{isa} extensions, which makes the \gls{fp} register file of the core \dbit{64} wide.
We add support for operations on FP16, FP16alt, and FP8, as well as \gls{simd} operations for all these formats, including FP32.
While we support the flexible cast-and-pack operations, this version of the core is not equipped with expanding \gls{fma} operations.

\subsubsection{Core Modifications}
We replace the core's \gls{fpu} with our design, extend the processor's decoder with the new operations and the load/store circuitry of the core to also allow for one-extending narrower \gls{fp} data for proper \acrshort{nan}-boxing.
These additional changes in core control circuitry are not timing-critical, and their cost is negligible concerning the rest of the core resources.

\subsubsection{FPU Configuration}

\begin{table}
  \centering
  \begin{threeparttable}
    \centering
    \caption{The configuration of the \gls{tpfpu} as implemented into the Ariane core. The FPU width is $\wfpu = \SI{64}{\bit}$.}
    \label{tab:config}
    \begin{tabularx}{\linewidth}{@{} Xllll @{}}
      \toprule
      \textbf{Format} &
      \multicolumn{4}{l}{\textbf{Implementation} (number of cycles, number of lanes)}\\
      &
      ADDMUL &
      DIVSQRT &
      COMP &
      CONV \\
      \midrule
      FP64    & parallel (4,1) & merged (21,1\tnote{*}) & parallel (1,1) & merged (2,2\tnote{*})\\
      FP32    & parallel (3,2) & merged (11,0) & parallel (1,2) & merged (2,0)\\
      FP16    & parallel (3,4) & merged (7,0)  & parallel (1,4) & merged (2,2\tnote{*})\\
      FP16alt & parallel (3,4) & merged (6,0)  & parallel (1,4) & merged (2,0)\\
      FP8     & parallel (2,8) & merged (4,0)  & parallel (1,8) & merged (2,4)\\
      \bottomrule
    \end{tabularx}
    \begin{tablenotes}
      \item[*] Merged lane with support for all formats of equal width and narrower.
    \end{tablenotes}
  \end{threeparttable}
  \vspace{-0.5cm}
\end{table}

We configure the \gls{tpfpu} to include the aforementioned formats and add format-specific pipeline depths as shown in \tabref{tab:config}.
The number of pipeline registers is set so that the processor core can achieve a clock frequency of roughly \SI{1}{\giga\hertz}.
$\wfpu$ is set to the \gls{fp} register file width of \SI{64}{\bit}, hence there are no \gls{simd} vectors for the FP64 format.

We choose a parallel implementation of the ADDMUL block to vary the latency of operations on different formats and not incur unnecessary energy and latency overheads for narrow \gls{fp} formats.
Latency is format-dependent for DIVSQRT due to the iterative nature of the divider hardware used and not available on \gls{simd} data to conserve area.
In addition to a constant three cycles for pre- and post-processing, three mantissa bits are produced every clock cycle.
Divisions take between 4 (FP8) and 21 (FP64) cycles, which is acceptable due to the relative rarity of divide and square-root operations in performance-optimized code.
Conversions are again implemented using a merged slice, where two lanes are \SI{64}{\bit} wide for cast-and-pack operations using two FP64 values.
Additionally, there are two and four \dbit{16} and \dbit{8} lanes, respectively, to cover all possible conversions.

\section{Implementation Results}
\label{sec:results}

\subsection{PULPissimo: RI5CY with Transprecision FPU}
\label{sec:riscyimpl}
In order to benchmark applications on the TP-enabled RI5CY core, we perform a full place \& route implementation of a platform containing the core.
This section presents the implementation results, while \secref{sec:programming} shows an application case study on the implemented design.

\subsubsection{Implementation}

\begin{figure}
  \centering
  \includegraphics[width=0.8\linewidth]{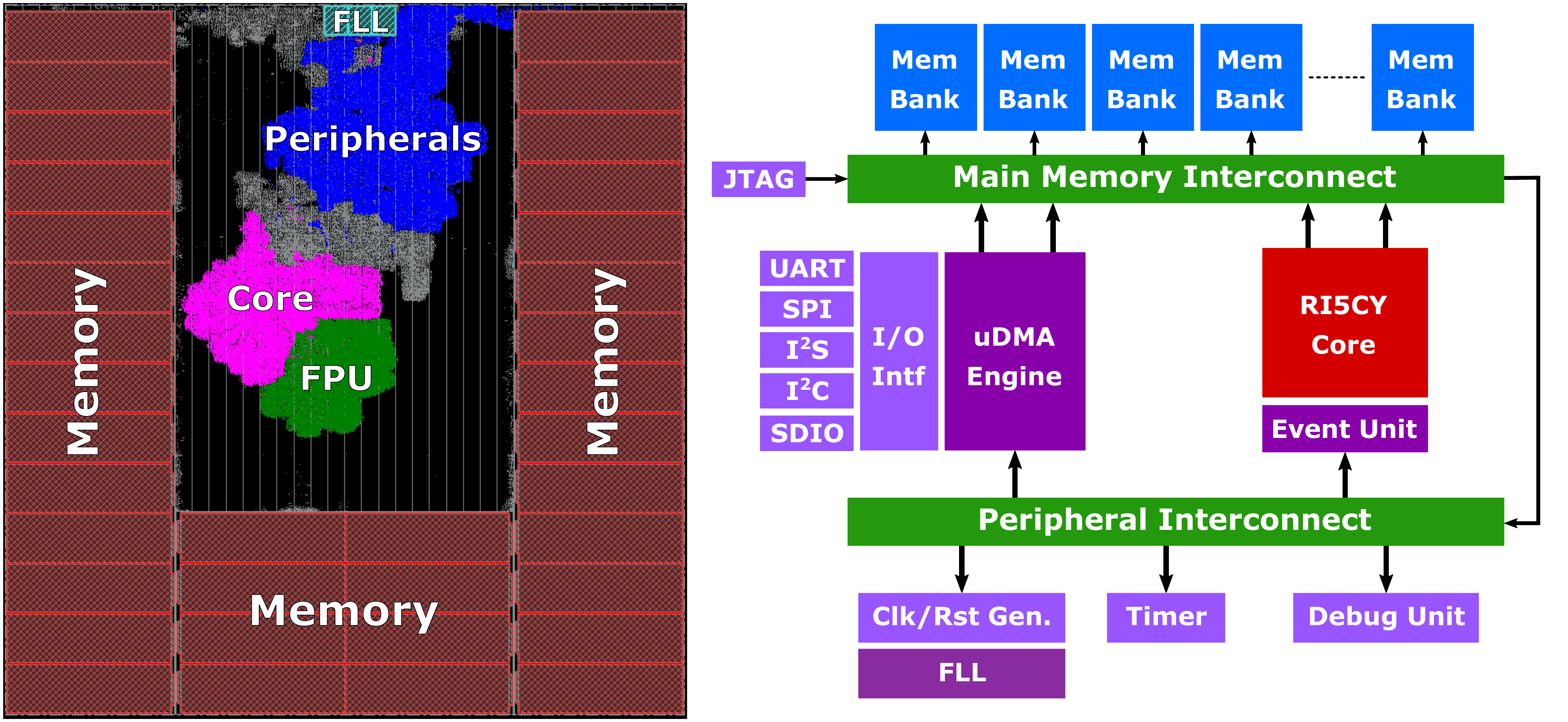}
  \caption{Floorplan and block diagram of the PULPissimo SoC (without pad frame) using RI5CY Core with TP-FPU.}
  \vspace{-0.5cm}
  \label{fig:pulpissimp_fp}
\end{figure}

We make use of PULPissimo\footnote{\url{https://github.com/pulp-platform/pulpissimo}} to implement a complete system.
PULPissimo is a single-core SoC platform based on the RI5CY core, including \SI{512}{\kilo\byte} of memory as well as many standard peripherals such as UART, I\textsuperscript{2}C, and SPI.
We use our extended RI5CY core as described in \secref{sec:implriscy}, including the single-cycle \gls{tpfpu} configuration shown in \tabref{tab:config_riscy}.

The system has been fully synthesized, placed, and routed in \gf technology, a \SI{22}{\nano\meter} FD-SOI node, using a low-threshold 8-track cell library at low voltage.
The resulting layout of the entire SoC (sans I/O pads) is shown in \figref{fig:pulpissimp_fp}.
Synthesis and place \& route were performed using Synopsys Design Compiler and Cadence Innovus, respectively, using worst-case low-voltage constraints (SSG, \SI{0.59}{\volt}, \SI{125}{\celsius}), targeting \SI{150}{\mega\hertz}.
Under nominal low-voltage conditions (TT, \SI{0.65}{\volt}, \SI{25}{\celsius}), the system runs at \SI{370}{\mega\hertz}.
The critical path of the design is between the memories and the core, involving the SoC interconnect.

\subsubsection{Impact of the \gls{tpfpu}}

\begin{figure}
  \centering
  \includegraphics[width=\linewidth]{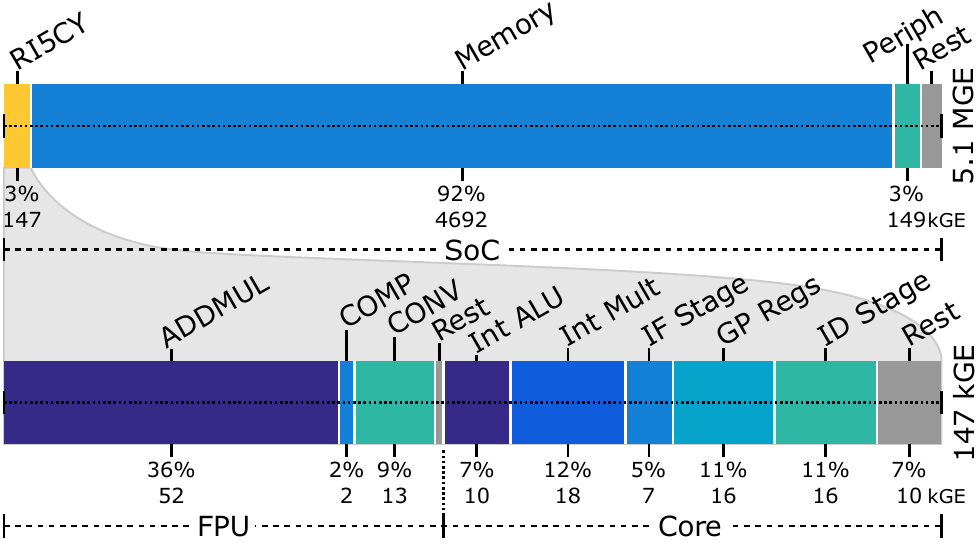}
  \vspace{-0.4cm}
  \caption{Area distribution of the PULPissimo SoC and the RI5CY core (in \si{\kilo\gateequivalent}, $\SI{1}{\gateequivalent}\approx\SI{0.199}{\micro\meter\squared}$).
  }
  \label{fig:area_riscy}
  \vspace{-0.5cm}
\end{figure}

The total area of the RI5CY core with \gls{tpfpu} is \SI{147}{\kilo\gateequivalent}, of which the \gls{fpu} occupies \SI{69}{\kilo\gateequivalent} ($47\%$), while the entire PULPissimo system including memories is \SI{5.1}{\mega\gateequivalent}, see \figref{fig:area_riscy}.
The ADDMUL block hosting the merged multi-format \gls{fma} units for all formats occupies 76\% of the \gls{fpu} area, while the COMP and CONV blocks use 4\% and 18\%, respectively.

Compared to a standard RI5CY core with support for only FP32, area increases by 29\% and static energy by 37\%.
The higher increase in energy \wrt the added area stems from the \gls{fpu} utilizing relatively more short-gate cells than the rest of the processor due to areas of higher timing pressure.
On the system scale, the added area and static energy account for only 0.7\% and 0.9\%, respectively, due to the impact of memories (92\% and 96\% of system area and leakage, respectively).

From an energy-per-operation point of view, it is interesting to compare FP32 \gls{fma} instructions with the \dbit{32} integer \gls{mac} instructions available in RI5CY.
Under nominal low-voltage conditions at \SI{370}{\mega\hertz}, these \gls{fp} and integer instructions consume \SI{3.9}{\pico\joule} and \SI{1.0}{\pico\joule} in their respective execution units on average.
Considering the system-level energy consumption, operating on FP32 data averages \SI{22.2}{\pico\joule} per cycle while the integer variant would require \SI{21.2}{\pico\joule} for running a filtering kernel (see \secref{sec:programming}), achieving equal performance.
These small system-level differences in area and static and dynamic energy, imply that FP computations are affordable even in an MCU context.

\subsection{Kosmodrom: Ariane with Transprecision FPU}
\label{sec:kosmoresults}

We implement a full test system with the TP-enabled Ariane core in silicon and perform a detailed analysis of the per-operation energy efficiency of \gls{fp} instructions.

\subsubsection{Silicon Implementation}

\begin{figure}
    \centering
    \includegraphics[width=\linewidth,clip,trim=0cm 2.8cm 0cm 2.5cm]{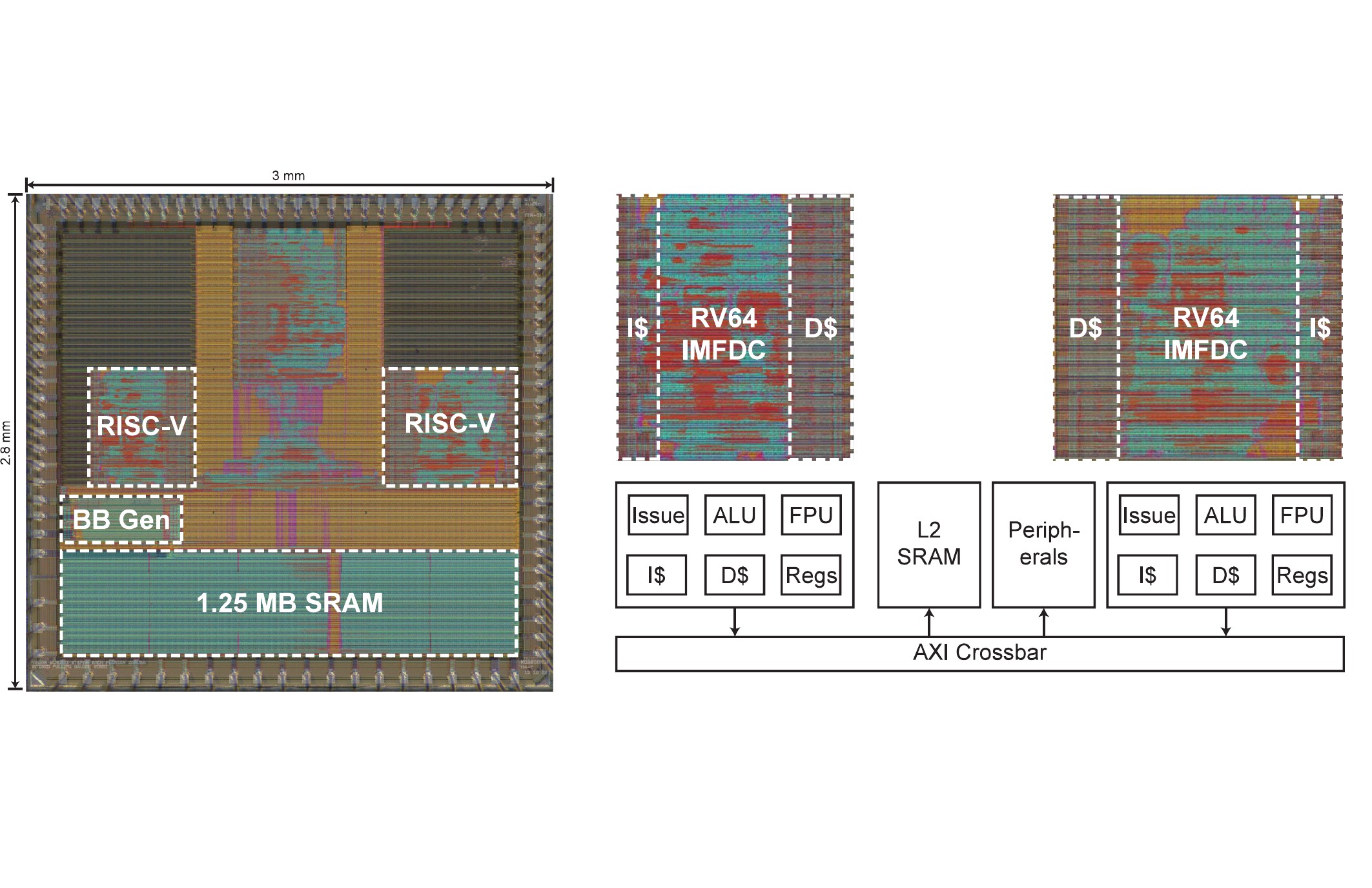}
    \vspace{-0.6cm}
    \caption{Die micrograph and block diagram of entire Kosmodrom chip showing the placement of the two Ariane core macros with TP-FPU.
    }
    \vspace{-0.5cm}
    \label{fig:floorplan}
\end{figure}

We implement a full test system in \gf technology called Kosmodrom \cite{zaruba2019floating}.
\figref{fig:floorplan} contains a silicon micrograph as well as an architectural overview of the main blocks in the design.
Two functionally identical Ariane cores with \gls{tpfpu} have been fabricated using different cell technologies and target frequencies, and share a \SI{1.25}{\mega\byte} L2 memory and common periphery like interrupt and debug infrastructure.
We support five \gls{fp} formats with dedicated datapaths for each one, leveraging the format-specific latencies shown in \tabref{tab:config}.

Synthesized using Synopsys Design Compiler, the faster, higher performance core uses a low threshold eight-track cell-library while the slower, low-power core, features a 7.5-track library.
For the subsequent performance and efficiency analysis, we will solely focus on the high-performance core as the cores can be individually clocked and powered.
In synthesis, a \SI{1}{\GHz} worst-case constraint (SSG, \SI{0.72}{\volt}, \SI{125}{\celsius}) was set.
We use automated clock gate insertion extensively during synthesis ($>96\%$ of \gls{fpu} registers are gated).
Ungated registers comprise only the handshaking tokens and the finite-state machine controlling division and square root.
The locations of pipeline registers in the entire \gls{fpu} were optimized using the register retiming functionality of the synthesis tool.

Placement and routing are done in Cadence Innovus with a \SI{1}{\GHz} constraint in a Multi-Mode Multi-Corner flow that includes all temperature and mask misalignment corners, eight in total.
The finalized backend design reaches \SI{0.96}{\GHz} under worst-case conditions (SSG, \SI{0.72}{\volt}, \SI{\pm0.8}{\volt} bias, \SI{-40/125}{\celsius}), \SI{1.29}{\GHz} under nominal conditions (TT, \SI{0.8}{\volt}, \SI{\pm0.45}{\volt} bias, \SI{25}{\celsius}), and \SI{1.76}{\GHz} assuming best case conditions (FFG, \SI{0.88}{\volt}, \SI{0}{\volt} bias, \SI{-40/125}{\celsius}).

We have also performed a substantial exploration of logic cell mixes (threshold voltage and transistor length) to maximize energy efficiency.
The design contains 74\% LVT and 26\% SLVT cells; and 86\% \SI{28}{\nano\meter}, 8\% \SI{24}{\nano\meter}, and 6\% \SI{20}{\nano\meter} transistors.

\subsubsection{Static Impact of the \gls{tpfpu}}

\begin{figure}
    \centering
    \includegraphics[width=\linewidth]{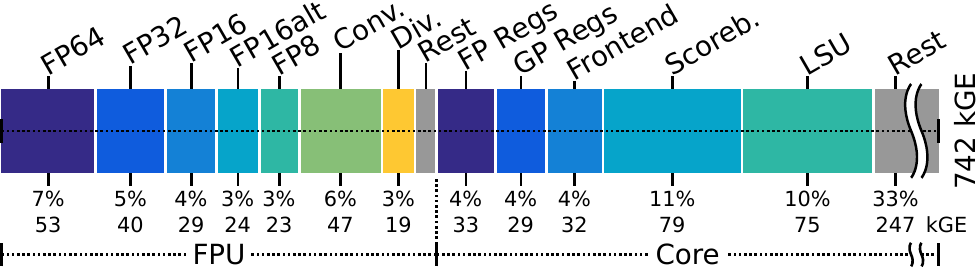}
    \vspace{-0.4cm}
    \caption{Area distribution of the entire Ariane \riscv{} core, excluding cache memories (in \si{\kilo\gateequivalent}, $\SI{1}{\gateequivalent}\approx\SI{0.199}{\micro\meter\squared}$).
    }
    \label{fig:area}
    \vspace{-0.5cm}
\end{figure}

The total area of the Ariane core with \gls{tpfpu} is \SI{742}{\kilo\gateequivalent} (\SI{1.4}{\mega\gateequivalent} including caches).
The area breakdown is shown in \figref{fig:area}.
The total size of the FPU is \SI{247}{\kilo\gateequivalent}, of which \SI{160}{\kilo\gateequivalent} make up the various \gls{fma} units, \SI{13}{\kilo\gateequivalent} are comparison and bit manipulation circuitry, \SI{19}{\kilo\gateequivalent} for the iterative divider and square root unit, and \SI{47}{\kilo\gateequivalent} are spent on the conversion units.
Compared to a complete Ariane core (including caches) with support for only scalar FP32 and FP64 (`F' and `D' extensions), area and static energy are increased by 9.3\% and 11.1\%, respectively.
The added area and energy cost in the processor are moderate, considering that \gls{fp} operations on three new formats were added, along with \gls{simd} support, which improves \gls{fp} operation throughput by up to 8$\times$ when using FP8 vectors.

\subsubsection{Silicon Measurements}
\paragraph{Evaluation Methodology}

We extract a detailed breakdown of energy consumption within the \gls{fpu} by stressing individual operations using synthetic applications on Ariane.
Each instruction is fed with randomly distributed normal \gls{fp} values constrained such that the operations do not encounter overflow, creating a worst-case scenario for power dissipation by providing high switching activity inside the datapath of the \gls{tpfpu}.
Measurements are taken with full pipelines and the \gls{fpu} operating at peak performance to provide a fair comparison.

Silicon measurements of the core and memory power consumption are done with the processor and \gls{fpu} performing a matrix-matrix multiplication.
Using a calibrated post-place-and-route simulation with full hierarchical visibility allows us to determine the relative energy cost contribution of individual hardware blocks.
Post-layout power simulations are performed using typical corner libraries at nominal conditions (TT, VDD = \SI{0.8}{\volt}, \SI{25}{\celsius}).
Silicon measurements are performed under unbiased nominal (\SI{0.8}{\volt}, \SI{0}{\volt} bias, \SI{25}{\celsius}) conditions where \SI{923}{\mega\hertz} are reached, unless noted otherwise.
The impact of voltage scaling on the performance and energy efficiency are obtained through measurements of the manufactured silicon.

\paragraph{FPU Instruction Energy Efficiency and Performance}

\begin{figure}
    \centering
    \includegraphics[width=\linewidth]{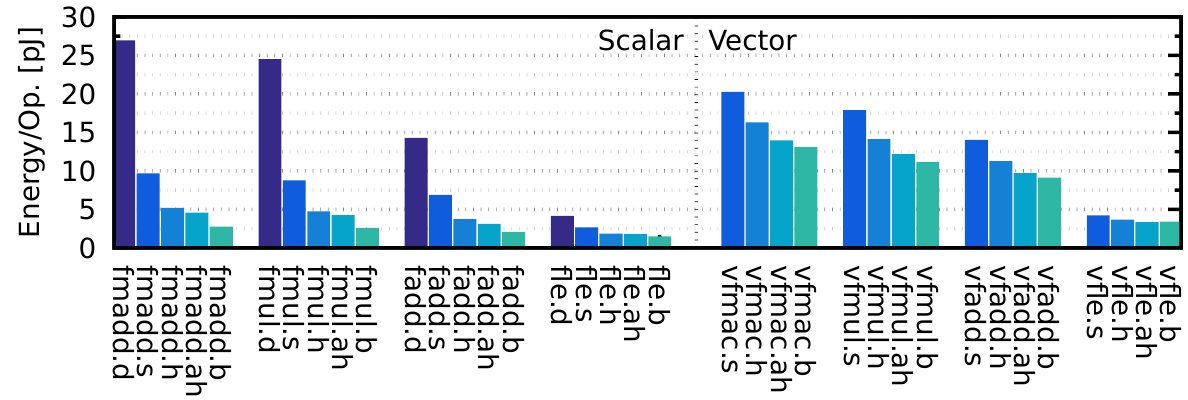}
    \includegraphics[width=\linewidth]{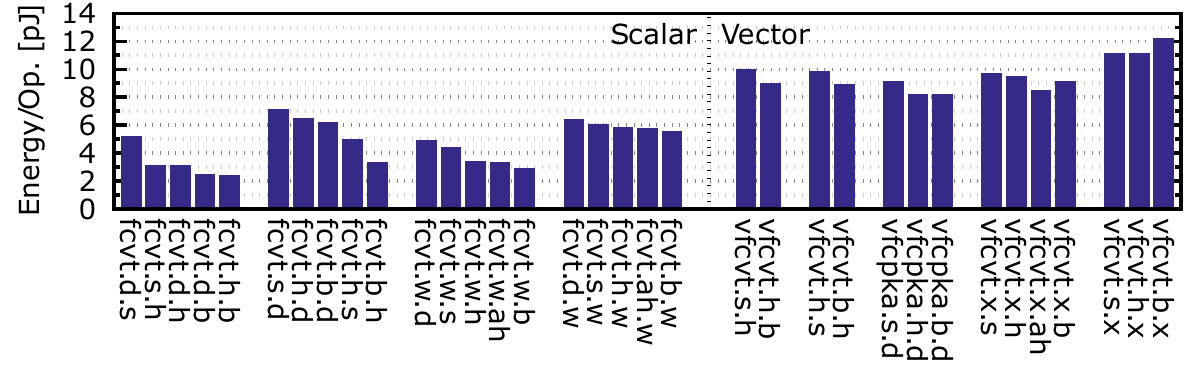}
    \vspace{-0.5cm}
    \caption{FPU energy cost per instruction for the fully pipelined scalar operations (top left), vectorial operations (top right), grouped by FMA, multiply, add, and comparison; and scalar conversion operations (bottom left) and vectorial conversion operations (bottom right).}
    \label{fig:fpenergy}
    \vspace{-0.5cm}
\end{figure}

The top of \figref{fig:fpenergy} shows the average \emph{per-instruction}\footnote{One \gls{fpu} instruction may perform multiple \si{\FLOP}s on multiple data items.} energy cost within the \gls{fpu} for arithmetic scalar operations.
Energy proportionality of smaller formats is especially pronounced in the ADDMUL block due to the high impact of the multiplier (first three groups of bars).
For example, the FP64 \gls{fma} \textit{fmadd.d} consumes \SI{26.7}{\pico\joule}, while performing the same operation on FP32 requires 65\% less energy.
Reducing the \gls{fp} format width further costs 48\%, 54\%, and 49\% of energy compared to the next larger format for FP16, FP16alt, and FP8, respectively.
Using FP16alt instead of FP16 consumes is energetically 12\% cheaper due to the smaller mantissa multiplier needed for FP16.
Similarly, reducing the \gls{fp} format width leads to relative energy gains compared to the next-larger format of $65\%$, $47\%$, $52\%$, $47\%$ for FP multiplication, $53\%$, $47\%$, $57\%$, $47\%$ for FP addition, and $38\%$, $34\%$, $35\%$, $22\%$ for FP comparisons using FP32, FP16, FP16alt, and FP8, respectively.
As such, scalar operations on smaller formats are energetically at least directly proportionally advantageous.

Intuition would suggest that \gls{simd} instructions on all formats would require very similar amounts of energy due to the full utilization of the \dbit{64} datapath.
However, we find that vectorial operations are also progressively energy-proportional, amplifying the energy savings even further.
Starting from an energy cost of \SI{20.0}{\pico\joule} for an FP32 \gls{simd} \gls{fma} \textit{vfmac.s}, the \emph{per-instruction} energy gains to the next-larger format for FP16, FP16alt, and FP8 are $20\%$, $31\%$, and $20\%$. Similarly, they are $21\%$, $32\%$, and $21\%$ for multiplication, $20\%$, $31\%$, and $19\%$ for addition, and $14\%$, $23\%$, and $8\%$ for comparisons.
Despite the full datapath utilization, packed operations using more narrow \gls{fp} formats offer super-proportional energy gains while simultaneously increasing the throughput per instruction.
This favorable scaling is owed to the separation in execution units for the individual formats where idle slices are clock-gated, which would be harder to attain using a conventional shared-datapath approach.
By accounting for the increased throughput, the \emph{per-datum} energy gains to the next larger format become $60\%$, $66\%$, and $58\%$, for the \gls{simd} \gls{fma}, which is better than direct proportionality.

Conversion instructions that share a merged slice for all formats in the CONV block of the architecture are an example of less pronounced energy scaling.
The bottom of \figref{fig:fpenergy} shows the average \emph{per-instruction} energy consumption of conversions on scalars and vectors.
Energy consumption of the instructions is influenced by both the source and destination formats in use.

For scalar \gls{fp}-\gls{fp} casts, converting to a larger format is energetically cheaper as only part of the input datapath toggles and most of the output mantissa is padded with constant zeroes.
Casts to a smaller format are more expensive as the wide input value causes dynamic switching within the conversion unit to produce the output.
To contrast with the scaling results obtained above, we compare conversions where both the input and output formats are halved, such as \texttt{fcvt.s.d}, \texttt{fcvt.h.s}, and \texttt{fcvt.b.h}.
Starting from \SI{7.0}{\pico\joule} for the FP64/FP32 cast, we find a reduction in energy of merely $30\%$ and $35\%$ when halving the format widths.
Compared to the energy scaling results from above, the scaling is worse due to the use of one merged unit where unused portions of the datapath are much harder to turn off.

For \gls{simd} vectors, the effect of \emph{per-instruction} energy proportionality is visible again, going from the FP32/FP16 cast to the FP16/FP8 cast is $9.5\%$ cheaper.
While not as significant as for vectorial \gls{fma}, this gain is due to the additional vector lanes for casting small formats being narrower and supporting fewer formats.

The flexible cast-and-pack instructions allow the conversion of two FP64 values and pack them into two elements of the destination vector for only roughly $30\%$ more energy than performing one scalar conversion from FP64 to the target format.
It should be noted that two scalar casts and additional packing operation, which is not directly available in the \gls{isa}, would be required without this functionality.

Measuring scalar \gls{fp}-integer conversions where the integer width is fixed also shows the relatively small relative gains, up to only $25\%$ for FP16alt/int32 vs. FP32/int32 conversions, much worse than direct proportionality.
Vectorial \gls{fp}-integer casts operate on integers of the same width as the \gls{fp} format.
Here, the impact of sharing vectorial lanes with other formats can make \gls{simd} cast \emph{instructions} on many narrow values cost more energy than on the larger formats, such as for the FP8/int8 cast, diminishing \emph{per-datum} energy scaling compared to the parallel slices.

Under nominal conditions, our \gls{tpfpu} thus achieves scalar \gls{fma} in \SIrange{2.5}{26.7}{\pico\joule}, \gls{simd} \gls{fma} in \SIrange{1.6}{10.0}{\pico\joule} per data item, over our supported formats.
\gls{fp}-\gls{fp} casts cost \SIrange{7.0}{26.7}{\pico\joule} for scalar, and \SIrange{2.2}{4.9}{\pico\joule} for vectorial data, respectively.
Our approach of dividing the unit into parallel slices has proven to be effective at achieving high energy proportionality on scalar and \gls{simd} data.
At the silicon's measured nominal frequency of \SI{923}{\MHz} this corresponds to a performance and energy efficiency of \SIrange{1.85}{14.83}{\giga\FLOPs} and \SIrange{75}{1245}{\giga\FLOPsW} for the \gls{fma} across formats.

\paragraph{Impact of Voltage Scaling}
\label{sec:vscaling}

\begin{figure}
    \centering
    \includegraphics[width=\linewidth]{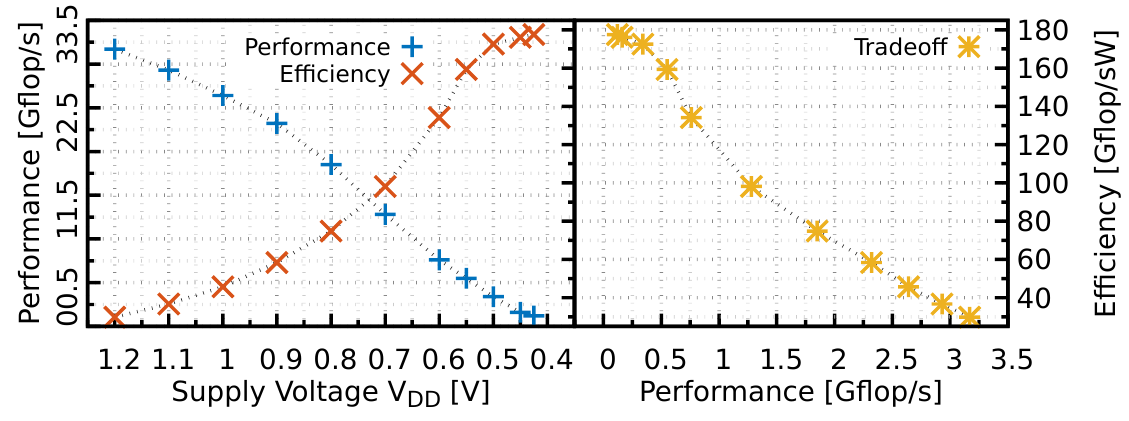}
    \caption{Compute performance and energy efficiency of FP64 \gls{fma} versus supply voltage (left), trade-off between compute performance and energy efficiency, achieved by adjusting supply voltage and operating frequency. Measured on manufactured silicon.}
    \label{fig:vddsweeps}
\end{figure}

\figref{fig:vddsweeps} shows the impact of voltage and frequency scaling on the manufactured silicon.
We measure the highest possible frequency and corresponding power consumption for supply voltages between \SI{0.425}{\volt} and \SI{1.2}{\volt}.
We observe peak compute and efficiency numbers of \SI{3.17}{\GFLOPs} and \SI{178}{\GFLOPsW} for FP64, \SI{6.33}{\GFLOPs} and \SI{473}{\GFLOPsW} for FP32, \SI{12.67}{\GFLOPs} and \SI{1.18}{\TFLOPsW} for FP16, \SI{12.67}{\GFLOPs} and \SI{1.38}{\TFLOPsW} for FP16alt, and \SI{25.33}{\GFLOPs} and \SI{2.95}{\TFLOPsW} for FP8.


\paragraph{Core-Level Energy Efficiency}

\begin{figure}
  \centering
  \includegraphics[width=\linewidth]{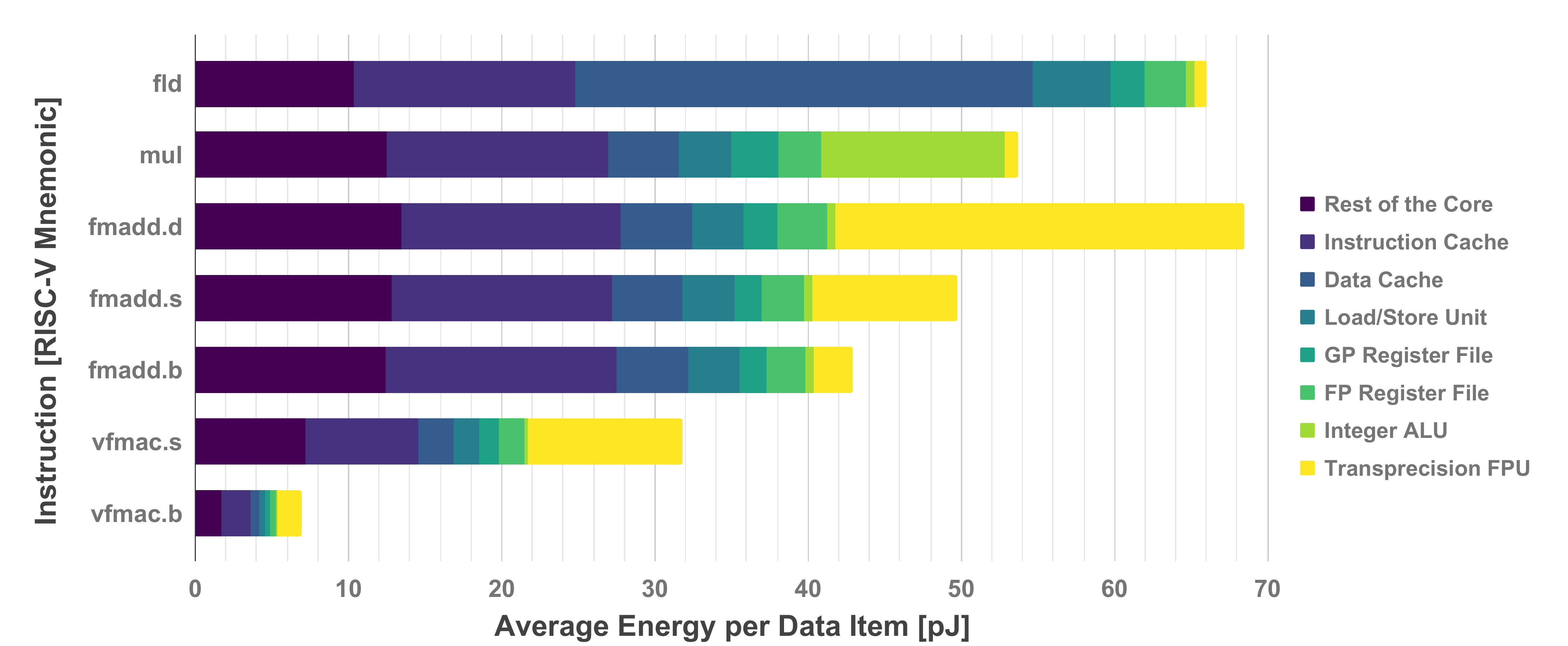}
  \vspace{-0.8cm}
  \caption{Energy cost per data item of operations in the entire Ariane core.}
  \label{fig:core_pwr}
  \vspace{-0.5cm}
\end{figure}

As the \gls{tpfpu} is merely one part of the entire processor system, we now briefly consider the energy spent during operations within the entire core.
\figref{fig:core_pwr} shows the per-data energy consumption of the processor blocks performing various operations in the Ariane core.
During an FP64 \gls{fma} - energetically the most expensive \gls{fp} operation - the \gls{fpu} accounts for $39\%$ of the total Ariane core energy, with energy consumption of memory operations being comparable with that of the FP64 \gls{fma}.
Although thanks to formidable energy proportionality, the FP8 FMA consumes $10.5\times$ less \gls{fpu} energy than the same operation on FP64, overall core energy consumption is decreased by only $38\%$.
While for small and embedded applications scalar \gls{fpu}-level energy savings might be sufficient, they are not enough to bring maximum savings in energy efficiency through transprecision in application-class cores such as Ariane due to the relatively large core-side overheads.

Employing \gls{simd} vectorization strongly mitigates this core overhead's impact on the energy cost per item.
For example, the FP8 \gls{fma} requires another $6.2\times$ less total core energy when executed as part of a vectorial FMA.

\subsection{Performance and Programming of Transprecision Kernels}
\label{sec:programming}

To visualize some challenges and benefits of transprecision applications, we showcase a multi-format application kernel running on the transprecision-enabled RI5CY core.
Furthermore, we touch on the considerations to make when programming for transprecision-enabled platforms.

\subsubsection{Transprecision Application Case Study}

We consider the accumulation of element-wise products of two input streams, commonly found in many applications such as signal processing or SVM.

\begin{figure}
  \centering
  \includegraphics[width=\linewidth]{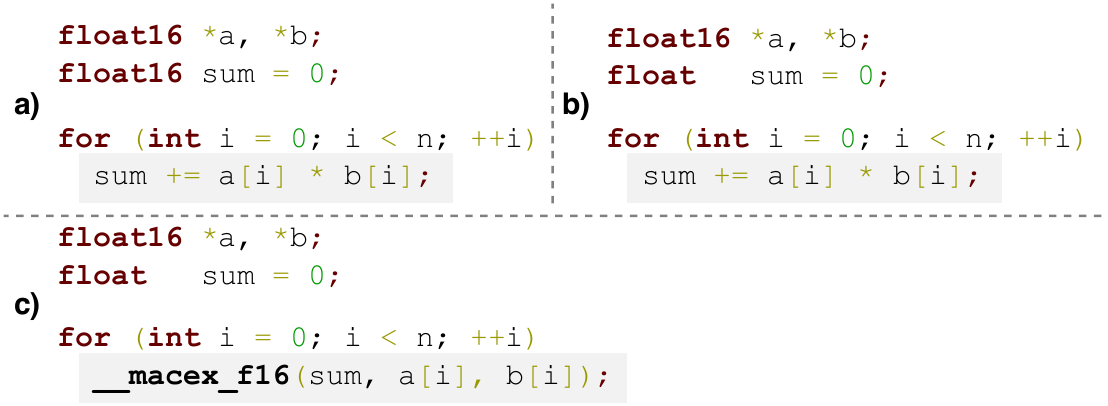}
   \vspace{-0.5cm}
  \caption{Accumulation of element-wise products from two input streams \texttt{a} and \texttt{b}. Inputs are in FP16, the result is accumulated using FP16 in a), and using FP32 otherwise. Code c) uses compiler intrinsic functions to invoke transprecision instructions.}
  \vspace{-0.5cm}
  \label{fig:cloop}
\end{figure}

\begin{figure*}
  \centering
  \includegraphics[width=\textwidth]{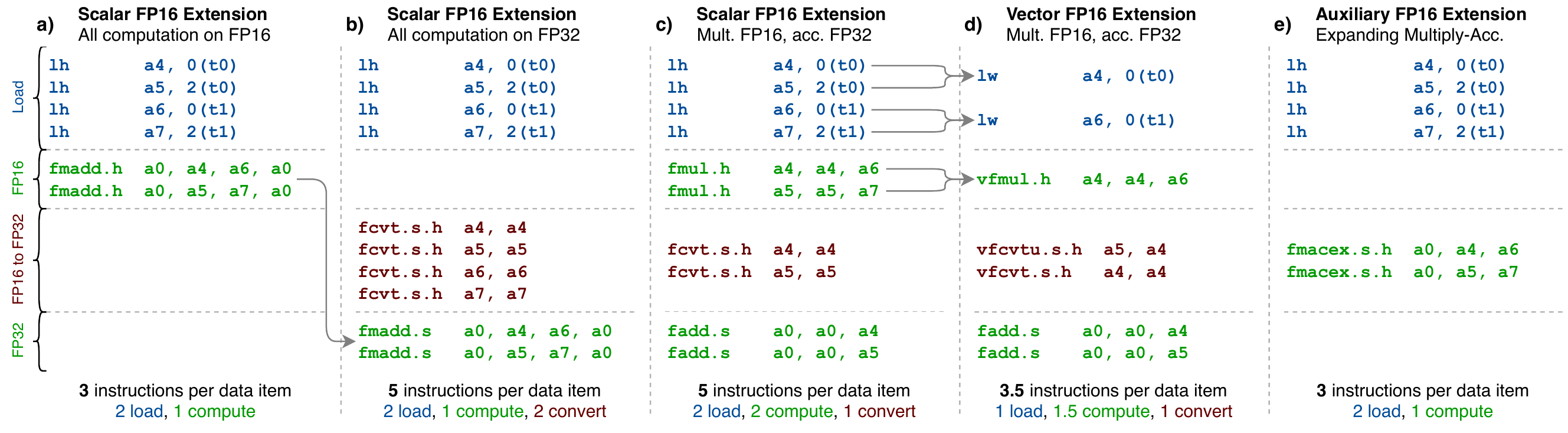}
  \caption{\riscv assembly implementations of two iterations of the loop body in \figref{fig:cloop} (grey). %
           }
  \label{fig:tpperf}
  \vspace{-0.5cm}
\end{figure*}

\begin{table}
  \centering
  \begin{threeparttable}
  \centering
  \caption{Application metrics corresponding to assembly from \figref{fig:tpperf}}
  \label{tab:asmresults}
  \begin{tabularx}{\linewidth}{@{}Xk{2}k{1.1e-1}k{1.1e-1}k{1.2}k{1.2}@{}}
    \toprule
                                          & {\textbf{\# Bits}} & \multicolumn{2}{c}{\textbf{Rel. Error of Result vs.}} & \multicolumn{2}{c}{\textbf{Rel. Energy}} \\
                                          &  {correct}         & {Exact}     & {Exact FP16\tnote{*}} & {Core}    &  {System}\\
    \midrule
    \textbf{Exact Result}                 & 37                 & 0.0           &        &       & \\
    Cast to FP16                          & 12                 & 1.9e-4        & 0.0    &       & \\
    \midrule
    Result \ref{fig:tpperf}\,\textbf{a)}   & 9                  & 2.7e-3        & 2.9e-3 & 0.60  & 0.63  \\   
    Result \ref{fig:tpperf}\,\textbf{b)}   & 22                 & 2.0e-7        & 0.0    & 1.00  & 1.00  \\   
    Result \ref{fig:tpperf}\,\textbf{c)}   & 19                 & 1.6e-6        & 0.0    & 1.16  & 1.03  \\   
    Result \ref{fig:tpperf}\,\textbf{d)}   & 19                 & 1.6e-6        & 0.0    & 0.97  & 0.75  \\   
    Result \ref{fig:tpperf}\,\textbf{e)}   & 22                 & 2.0e-7        & 0.0    & 0.63  & 0.63  \\   
    \bottomrule
  \end{tabularx}
  \begin{tablenotes}
    \item[*] The final result is converted to FP16 and compared to the exact result converted to FP16
  \end{tablenotes}
  \end{threeparttable}
  \vspace{-0.5cm}
\end{table}

\paragraph{Approach}
\figref{fig:cloop} shows the C representation of the workload relevant for our evaluation.
The input streams reside in memory as FP16 values, and the accumulation result uses FP16 or FP32.
We use our transprecision \gls{isa} extensions to obtain the assembly in \figref{fig:tpperf} as follows:
  \figref{fig:tpperf}\,a) is the FP16-only workload in {\figref{fig:cloop}\,a)} requiring an ideal 3 instructions per input pair.
  \figref{fig:tpperf}\,b) performs all operations on FP32 to achieve the most precise results but requires casts in a total of 5 instructions.
  \figref{fig:tpperf}\,c) tries to save energy by performing the multiplication in FP16 to replace the FP32 \gls{fma} with additions.
  \figref{fig:tpperf}\,d) accelerates the FP16 portion of the previous code by using \gls{simd} in 3.5 instructions.
  \figref{fig:tpperf}\,e) makes use of expanding multi-format \gls{fma} instructions to combine computation and conversion in 3 instructions again.

The complete application repeats these actions over the entire input data using the zero-overhead hardware loops and post-incrementing load instructions available in RI5CY.
Further manual loop unrolling can only be used to hide instruction latency overheads due to the low data intensity of this workload.

\paragraph{Performance and Energy Results}
We collect the final result accuracy and energy use of these programs in \tabref{tab:asmresults}.
Energy results have been obtained from a post-layout simulation of the RI5CY + \gls{tpfpu} design presented in \secref{sec:riscyimpl}.

The accuracy of the result from \figref{fig:tpperf}a) is relatively low with \SI{9}{\bit} of precision correct (about three decimal digits), while the exact result of the operation would require \SI{37}{\bit} of precision.
Due to the accumulation of rounding errors, the result strays far from the possibly most accurate representation of the exact result in FP16 (\SI{12}{\bit} correct).
The code in \figref{fig:tpperf}\,b) offers \SI{22}{\bit} of precision but increases energy cost by 66\% and 59\% on core and system level, respectively, due to the increased execution time and higher per-instruction energy spent on FP32 operations.
\figref{fig:tpperf}\,c) suffers from decreased accuracy (\SI{19}{\bit}) and even requires 16\% more core energy (+3\% system energy) \wrt the FP32 code, as the FP16 multiplications are energetically much more expensive than the casts they replace.
Compared to the FP32 case, the use of \gls{simd} in \figref{fig:tpperf}\,d) reduces core energy by 3\% and total system energy by even 25\%.
In the core, the increased performance slightly outweighs the increased \gls{fpu} energy, where on the system level, the lower number of memory operations has a significant effect.
Using the expanding multiply-accumulate operations in \figref{fig:tpperf}\,e) offers the best of both worlds: same performance as the naïve FP16-only version, as well as the same precision as if performed entirely on FP32.
Converted to FP16, this yields a value $14.6\times$ more accurate than using FP16 only, reducing core and system power by 37\% compared to the FP32 case.
These results highlight the energy savings potential of transprecision computing when paired with flexible hardware implementations.

\subsubsection{Compiler Support}

We make the low-level transprecision instructions available as a set of compiler intrinsic functions that allow full use of the transprecision hardware by the compiler and programmer.
Scalar types and basic operations are transparently handled by the compiler through the usage of the appropriate types (\texttt{float16}, \texttt{float16alt}, \texttt{float8}), and operators (\texttt{+}, \texttt{*}, etc.).
Vectorial operations on custom \gls{fp} formats are inferred by using GCC vector extensions.
In fact, the compiler can generate programs such as in \figref{fig:tpperf}\,d) from the code in \figref{fig:cloop}\,b).

However, operations such as the \gls{fma} as well as optimized access and conversion patterns using \gls{simd} vectors often do not cleanly map to high-level programming language operators and semantics.
It is prevalent that performance-optimized \gls{fp} code requires low-level manual tuning to make full use of the available hardware, even in non-transprecision code.
We can and should make use of the non-inferrable operations such as cast-and-pack or expanding \gls{fma} through calls to intrinsics, as seen in \figref{fig:cloop}\,c), which can produce assembly \figref{fig:tpperf}\,e).
The benefits and limits of the compiler-based approach are further investigated in \cite{tagliavini2019extension}.

\section{Related Work}
\label{sec:relwork}


\subsection{\gls{simd} and Transprecision in Commercial \glspl{isa}}

Intel's x86-64 SSE/AVX extensions
offer very wide \gls{simd} operations (up to \SI{512}{\bit} in AVX-512) on FP32 and FP64.
They include an \gls{fp} dot-product instruction that operates on vectors of $2\times$ FP64 or $4\times$ FP32, respectively, producing a scalar result.
Currently, no non-standard \gls{fp} formats are supported, but future CPUs with the AVX-512 extension (Cooper Lake) will include the \emph{BF16} format (FP16alt) with support for cast-and-pack, as well as an expanding \gls{simd} dot-product on value pairs.

The ARM NEON extension
optionally supports FP16 and contains a separate register file for \gls{simd} operations that supports register fusion through differnt addressing views depending on the \gls{fp} format used.
The addressing mode is implicit in the use of formats within an instruction, enabling very consistent handling of multi-format (expanding, shrinking) operations that always operate on entire registers.
This approach contrasts with our \gls{isa} extension, which requires multiple encodings to slice input or output vectors during vectorial conversions.
A supplement to the ARM \gls{isa} is the Scalable Vector Extension (SVE)
, targeting high-performance \dbit{64} architectures only, providing scaling to vector lengths far beyond \SI{128}{\bit}.
SVE contains optional support for BF16 as a storage format, implicitly converting all BF16 input data to FP32 when used in computations, producing FP32 results.
Converting \gls{fp} data to BF16 for storage is also possible.

A new \gls{isa} extension for ARM M-class processors is called MVE
.
It reconfigures the FP register file to act as a bank of eight \dbit{128} vector registers, each divided into four 'beats' of \SI{32}{\bit}.
While vector instructions always operate on the entire vector register (fixed vector length of \SI{128}{\bit}), implementations are free to compute one, two, or all four beats per clock cycle -- essentially allowing serializing execution on lower-end hardware.
The floating-point variant of this \gls{isa} extension can operate on FP16 and FP32.
Multi-format operations are not supported.
An execution scheme in the spirit of MVE would apply to processors using our \gls{isa} extension with very little implementation overhead.
For a single vector instruction, emitting a sequence of four or two \gls{simd} \gls{fp} operations recreates the behavior of a single-beat or dual-beat system for a \gls{fp} register width of \SI{32}{bit} and \SI{64}{bit}, respecitvely.
Furthermore, MVE also supports predication on individual vector lanes, interleaving, and scatter-gather operations not available in our extension.

There exists a working draft for the \riscv `V' standard vector extension
.
The `V' extension adds a separate vector register file with Cray-style vector operation semantics and variable vector lengths.
Multiple registers can also be fused to increase the vector length per instruction effectively.
The standard vector extension includes widening and narrowing operations that fuse registers on one end of the operation, allowing consistent handling of data without the need for addressing individual register portions.
It supports FP16, FP32, FP64, and FP128, as well as widening \gls{fma} operations.
They operate in the same manner as our implementation of the \texttt{fmacex} operation, with the limitation that the target format must be exactly $2\times$ as wide as the source.
Furthermore, reduction operations for the inner sum of a vector exist.

\subsection{Open-Source Configurable FPU Blocks}
Most open-source \gls{fpu} designs implement a fixed implementation in a specific format, targetting a specific system or technology\footnote{\url{https://opencores.org/projects/fpu}}\footnote{\url{https://opencores.org/projects/fpu100}}, however there are some notable configurable works available.

For example, FloPoCo \cite{DinechinPasca2011-DaT} is a \gls{fp} function generator targeted mainly at \glspl{fpga} implementations, producing individual functions as VHDL entities.
\gls{fp} formats are freely configurable in terms of exponent and mantissa widths, the resulting hardware blocks are not \fpstd-compliant, however.
Namely, infinity and \gls{nan} values are not encoded in the \gls{fp} operands themselves, and subnormals are not supported as a trade-off for a slightly higher dynamic range present in FloPoCo \gls{fp} formats.
The \gls{fma} operation is not available.

Hardfloat\footnote{\url{https://github.com/ucb-bar/berkeley-hardfloat/}} on the other hand provides parametric \gls{fp} functions that are \fpstd compliant.
It is a collection of hardware modules written in Chisel with parametric \gls{fp} format and includes the \gls{fma} operator.
While Chisel is not widely adopted in commercial EDA tool flows, a generated standard Verilog version is also available.
Hardfloat internally operates on a non-standard recoded representation of \gls{fp} values.
However, the operations are carried out following \fpstd, and conversion blocks are provided to the standard interchange encoding, which is used in the \gls{tpfpu}.

Both of these works offer individual function blocks instead of fully-featured \glspl{fpu}.
However, thanks to the hierarchical architecture of the \gls{tpfpu}, it would be easily possible to replace its functional units with implementations from external libraries.

\subsection{\glspl{fpu} for \riscv}
Some vagueness exists in \fpstd concerning so-called \emph{implementation-defined behavior}, leading to problems with portability and reproducibility of \gls{fp} code across software and hardware platforms.
\gls{fp} behavior can be vastly different depending on both the processor model and compiler version used.
In order to avoid at least the hardware-related issues, \riscv specifies precisely how the open points of \fpstd are to be implemented, including the exact bit patterns of \gls{nan} results and when values are rounded.
As such, \glspl{fpu} intended for use in \riscv processors (such as this work), are usually consistent in their behavior.

The \glspl{fpu} used in the \riscv cores originating from UCB, Rocket and BOOM\cite{asanovic2016rocket, asanovic2015berkeley}, are based on Hardfloat components in specific configurations for \riscv.

Kaiser et al. \cite{kaiser2019development} have published a \riscv-specific implementation of the \gls{fma} operations on FP64 in the same \gf technology as this work.
We compare our \gls{tpfpu} to their implementation and others towards the end of this section.

\subsection{Novel Arithmetics / Transprecision \gls{fp} Accelerators}

Non-standard \gls{fp} systems are becoming ever more popular in recent years, driven mainly by the requirements of dominant machine learning algorithms.

For example, both the Google TPU \cite{jouppi2017datacenter} and \textsc{Nvidia}'s Tensor Cores \cite{amit2018extreme} provide high throughput of optmizied operations on reduced-precision \gls{fp} formats for fast neural network inference.
Both offer a custom format termed \emph{bfloat16}, using the same encoding as FP16alt in this work.
The latter furthermore supports FP16 as well as a new \dbit{19} \gls{fp} format called \emph{TensorFloat32}, which is formed from the 19 most significant bits of an FP32 input value, producing results in FP32.
However, both architectures omit certain features mandated in the standard such as denormal numbers or faithful rounding in pursuit of higher throughput and lower circuit area.
Such optimizations are orthogonal to this work and could be leveraged by adding further non-compliant modes to the functional units within our architecture.

Dedicated accelerators geared towards neural network training such as NTX \cite{schuiki2019ntx}, for example, employ non-standard multiply-accumulate circuits using fast internal fixed-point accumulation.
While not compliant to \fpstd, they can offer higher precision and dynamic range for accumulations.

\gls{fp}-related number systems are also being employed, such as for example \glspl{unum} \cite{gustafson2017end}, Posits \cite{gustafson2017beating} or \glspl{lns}.
\gls{unum}-based hardware implementations were proposed in \cite{glaser2018826, bocco2017hardware}.
Fast multplication and transcendental functions were implemented into a \riscv core in \cite{gautschi20164}.
While the focus of our \gls{tpfpu} is to provide \fpstd-like \gls{fp} capabilities, our work could be leveraged in several orthogonal ways to combine with these more exotic number systems.
For example, dedicated functional units for these formats could be included in the \gls{tpfpu} as new operation groups alongside the current \gls{fp} functions to accelerate specific workloads.
Furthermore, our functional unit implementations can be utilized as a starting point to implement some of these novel arithmetic functions.
The datapath necessary for posit arithmetic is very similar to a merged functional unit with a large number of possible input formats.
Lastly, the \gls{tpfpu} could be used as an architectural blueprint and filled with arbitrary functional units, leveraging our architecture's energy proportionality.

\subsection{Multi-Mode Arithmetic Blocks}

\begin{table}
\centering
\begin{threeparttable}
\centering
\caption{FMA comparison of this work (complete \gls{fpu}) with other standalone [multi-mode] architectures at nominal conditions.}
\label{tab:compare}

\renewcommand{\arraystretch}{0.9}
\begin{tabularx}{\linewidth}{@{}XXc k{2.2}k{2.2}k{3.2}r@{}}
\toprule
    \textbf{Format} &&
    \textbf{L/T\tnote{*}} &
    {\textbf{Perf.\tnote{\dag}}} &
    {\textbf{Energy}} &
    \multicolumn{2}{c}{\textbf{Energy Efficiency}} \\
    &&&
    {[\si{\GFLOPs}]} &
    {[\si{\pico\joule\per\FLOP}]} &
    {[\si{\GFLOPsW}]} &
    {rel.} \\

\midrule
    \multicolumn{7}{@{}l@{}}{\textbf{This Work},
        \SI{22}{\nano\meter},
        \SI{0.8}{\volt}\tnote{a},
        \SI{0.049}{\milli\meter\squared} (entire FPU),
        \SI{923}{\MHz}} \\[2pt]

    FP64    & scalar & 4/1 &  1.85 & 13.36 &   74.83 &  1.0$\times$ \\
    FP32    & scalar & 3/1 &  1.85 &  4.72 &  211.66 &  2.8$\times$ \\
    FP16    & scalar & 3/1 &  1.85 &  2.48 &  403.08 &  5.4$\times$ \\
    FP16alt & scalar & 3/1 &  1.85 &  2.18 &  458.56 &  6.1$\times$ \\
    FP8     & scalar & 3/1 &  1.85 &  1.27 &  786.30 & 10.5$\times$ \\
    FP32    & vector & 3/2 &  3.71 &  5.01 &  199.70 &  2.7$\times$ \\
    FP16    & vector & 3/4 &  7.42 &  2.01 &  497.67 &  6.7$\times$ \\
    FP16alt & vector & 3/4 &  7.42 &  1.72 &  581.96 &  7.8$\times$ \\
    FP8     & vector & 2/8 & 14.83 &  0.80 & 1244.78 & 16.6$\times$ \\

\midrule
    \multicolumn{7}{@{}l@{}}{\textbf{Kaiser \textit{et al.} \cite{kaiser2019development}},
        \SI{22}{\nano\meter},
        \SI{0.8}{\volt}\tnote{c},
        \SI{0.019}{\milli\meter\squared},
        \SI{1.8}{\GHz}} \\[2pt]

    FP64 & scalar & 3/1 & 3.60 & 26.40 & 37.88 \\

\midrule
    \multicolumn{7}{@{}l@{}}{\textbf{Manolopoulos \textit{et al.} \cite{manolopoulos2010efficient}},
        \SI{130}{\nano\meter},
        \SI{1.2}{\volt}\tnote{b},
        \SI{0.287}{\milli\meter\squared},
        \SI{291}{\MHz}} \\[2pt]

    FP64 & scalar & 3/1 & 0.58 & 60.53 & 16.52 & 1.0$\times$ \\
    FP32 & vector & 3/2 & 1.16 & 30.26 & 33.05 & 2.0$\times$ \\

\midrule
    \multicolumn{7}{@{}l@{}}{\textbf{Arunachalam \textit{et al.} \cite{arunachalam2018efficient}},
        \SI{130}{\nano\meter},
        \SI{1.2}{\volt}\tnote{b},
        \SI{0.149}{\milli\meter\squared},
        \SI{308}{\MHz}} \\[2pt]

    FP64 & scalar & 8/1 & 0.62 & 28.86 & 34.64 & 1.0$\times$ \\
    FP32 & vector & 8/2 & 1.23 & 14.43 & 69.29 & 2.0$\times$ \\

\midrule
    \multicolumn{7}{@{}l@{}}{\textbf{Zhang \textit{et al.} \cite{zhang2019efficient}},
        \SI{90}{\nano\meter},
        \SI{1}{\volt}\tnote{c},
        \SI{0.181}{\milli\meter\squared},
        \SI{667}{\MHz}} \\[2pt]

    FP64 & scalar & 3/1 & 1.33 & 32.85 &  30.44 & 1.0$\times$ \\
    FP32 & vector & 3/2 & 2.67 & 16.43 &  60.88 & 2.0$\times$ \\
    FP16 & vector & 3/4 & 5.33 &  8.21 & 121.76 & 4.0$\times$ \\

\midrule
    \multicolumn{7}{@{}l@{}}{\textbf{Kaul \textit{et al.} \cite{kaul20121}},
        \SI{32}{\nano\meter},
        \SI{1.05}{\volt}\tnote{a},
        \SI{0.045}{\milli\meter\squared},
        \SI{1.45}{\GHz}} \\[2pt]

    FP32              & scalar & 3/1 &  2.90 & 19.4  &  52.00 & 1.0$\times$ \\
    FP20\tnote{\ddag} & vector & 3/2 &  5.80 & 10.34 &  96.67 & 1.9$\times$ \\
    FP14\tnote{\ddag} & vector & 3/4 & 11.60 &  6.2  & 161.11 & 3.1$\times$ \\

\midrule
        \multicolumn{7}{@{}l@{}}{\textbf{Pu \textit{et al.} \cite{pu2016fpmax}},
            \SI{28}{\nano\meter},
            \SI{0.8}{\volt}\tnote{a},
            \SI[parse-numbers=false]{0.024\tnote{\S}/0.018\tnote{\textbar}}{\milli\meter\squared},
            \SI[parse-numbers=false]{910\tnote{\S}/1360\tnote{\textbar}}{\MHz}}\\[2pt]

        FP64 & scalar & 6/1 & 1.82 & 45.05 &  43.70 & 1.0$\times$  \\
        FP32 & scalar & 6/1 & 2.72 & 18.38 & 110.00 & 2.5$\times$  \\



\bottomrule
\end{tabularx}

\begin{tablenotes}
    \item[*] Latency [\si{cycle}] / Throughput [\si{operation \per cycle}]
    \item[\dag] 1 \gls{fma} = \SI{2}{{\FLOP}s}
    \item[a] Silicon measurements
    \item[b] Post-layout results
    \item[c] Post-synthesis results
    \item[\ddag] FP20~=~FP32 using only \SI{12}{\bit} of precision, FP14~=~FP32 using only \SI{6}{\bit} of precision
    \item[\S] FP64 FMA design
    \item[\textbar] FP32 CMA design
\end{tablenotes}
\vspace{-0.3cm}
\end{threeparttable}
\end{table}

To our knowledge, no fully-featured \glspl{tpfpu} with support for multiple formats have been published so far.
However, multi-mode \gls{fma} architectures have been proposed recently, usually featuring computations on two or three \gls{fp} formats \cite{manolopoulos2010efficient, arunachalam2018efficient, zhang2019efficient, kaul20121}, or a combination of interger and \gls{fp} support \cite{bruintjes2012sabrewing}.
Table~\ref{tab:compare} compares the proposed architectures with our implementation under nominal conditions.
Note that results for our work measure the entire \gls{tpfpu} energy while performing the \gls{fma} operation, not just the \gls{fma} block in isolation as in the related works.

The \riscv-compatible \gls{fma} unit from \cite{kaiser2019development} supports only FP64 with no support for FP32 even though required by \riscv.
Synthesized in the same \SI{22}{\nm} technology as our implementation, it achieves a $49\%$ lower energy efficiency than our \gls{fpu} performing the same \gls{fma} operation.

The architectures in \cite{manolopoulos2010efficient, arunachalam2018efficient, zhang2019efficient} focus heavily on hardware sharing inside the \gls{fma} datapath which forces all formats to use the same latency, no support for scalars in smaller formats, as well as lack of substantial energy proportionality.
These architectures only achieve directly proportional energy cost, while our energy efficiency gains become subsequently better with smaller formats -- reaching $16.6\times$ lower energy for operations on FP8 \wrt FP64 (width reduction of $8\times$).
By using the voltage scaling knob, this efficiency gain can again be increased by $2.3\times$, allowing for an over-proportional benefit to using the narrow \gls{fp} formats in our implementation rather than a simple 2:1 trade-off.

The \gls{fma} implementation in \cite{kaul20121} uses a vectorization scheme where the FP32 mantissa datapath is divided by 2 or 4 employing very fine-grained gating techniques while keeping the exponent at a constant \dbit{8} width.
This architecture's intended use is to attempt a bulk of FP32 computations at $4\times$ throughput using the packed narrow datapath, costing $3.1\times$ less energy.
By tracking uncertainty, imprecise results are recomputed using the $2\times$ reduced datapath before reverting the operation in full FP32.
As such, the intermediate formats used in this unit do not correspond to any standard \fpstd formats.

FPMax \cite{pu2016fpmax} features separate implementations of the \gls{fma} operation for FP32 and FP64 without any datapath sharing, targeting high-speed \glspl{asic}.
Comparing the energy cost of their two most efficient instances (using different internal architectures) yields energy proportionality slightly lower than our full \gls{fpu} implementation.
This result further compounds the value in offering separate datapaths for different formats on the scale of the entire \gls{fpu}.
It prompts us to explore the suitability of specific \gls{fma} architectures for different formats in the future.

\subsection{Other uses of our \gls{tpfpu}}

The open-source nature of FPnew, as well as the fact that it is written in synthesizable SystemVerilog, lower the burden of implementing \gls{fp} functionality into new systems without the need for extra IP licenses or changes to standard design flow.
FPnew or subcomponents of it have found use under the hood of some recent works.

Ara\cite{cavalcante2019ara}, a scalable \riscv vector processor implementing a draft version of the `V' extension, makes use of FPnew instances to perform efficient matrix operations on 16 \dbit{64} vector lanes.

Snitch\cite{zaruba2020snitch} is a tiny pseudo dual-issue \riscv processor paired with a powerful double-precision \gls{fpu}.
It employs an instance of FPnew to provide FP64 and \gls{simd} FP32 compute capabilities.

GAP9\footnote{\url{https://greenwaves-technologies.com/gap9iotapplicationprocessor}}, a commercial \gls{iot} application processor announced by GreenWaves Technologies contains FPnew to enable sub-\dbit{32} transprecision \gls{fp} computation.

Much of the \gls{fp} hardware found in the European Processor Initiative (EPI) project\cite{kovavc2019european} is based on the open-source FPnew design.

\section{Conclusion}

We have presented FPnew, a configurable open-source \acrlong{tpfpu} capable of supporting arbitrary \gls{fp} formats.
It offers \gls{fp} arithmetic and efficient casting and packing operations, in both scalar and \gls{simd}-vectorized variants, with high energy efficiency and proportionality.
We implemented the \gls{tpfpu} into an embedded \riscv processor core to show the potential of transprecision computing, using our transprecision \riscv \gls{isa} extension.
In our case study, we achieve FP32 precision without incurring any performance overhead compared to an optimal scalar FP16 baseline, reducing system energy by 34\% \wrt the FP32 implementation.
Furthermore, we implement the unit as part of a \riscv application-class core into the first full \gls{tpfpu} silicon implementation with support for five \gls{fp} formats, in \gf.
Adaptive voltage and frequency scaling allows for energy efficiencies up to \SI{2.95}{\TFLOPsW} and compute performance up to \SI{25.33}{\GFLOPs} for 8$\times$FP8 \gls{simd} operation.
The cost in the additional area ($9.3\%$) and static energy ($11.1\%$) in the processor are tolerable in light of the significant gains in performance and efficiency possible with the \gls{tpfpu}.

Our design achieves better energy efficiency scaling than other multi-mode \gls{fma} designs thanks to the parallel datapaths approach taken in our architecture.
Thanks to its open nature, FPnew can be utilized in many different application scenarios, having found use both in embedded \gls{iot} applications and high-performance vector processing accelerators.



%





\ifCLASSOPTIONcaptionsoff
  \newpage
\fi



\bibliographystyle{IEEEtran}
\bibliography{IEEEabrv,refs}
%



%

\vfill

\begin{IEEEbiography}[{\includegraphics[width=1in,height=1.25in,clip,keepaspectratio]{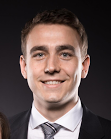}}]{Stefan Mach}
  received his B.Sc. and M.Sc. degree from the Swiss Federal Institute of Technology Zurich (ETHZ), Switzerland, where he is currently pursuing a Ph.D. degree. Since 2017, he has been a research assistant with the Integrated Systems Laboratory at ETHZ. His research interests include transprecision computing, computer arithmetics and energy-efficient processor architectures.
\end{IEEEbiography}

\begin{IEEEbiography}[{\includegraphics[width=1in,height=1.25in,clip,keepaspectratio]{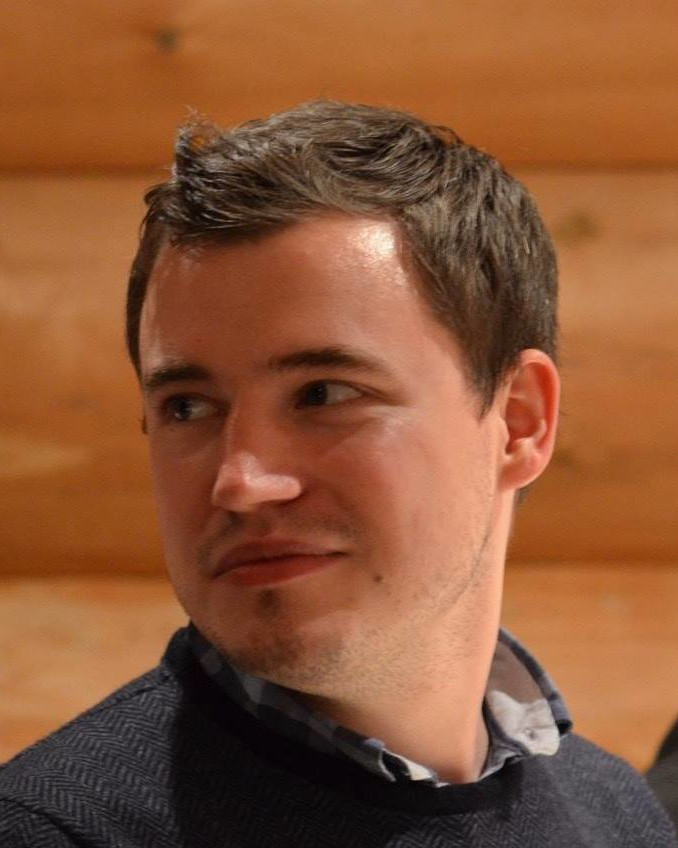}}]{Fabian Schuiki}
  received the B.Sc. and M.Sc. degree in electrical engineering from ETH Zürich, in 2014 and 2016, respectively. He is currently pursuing a Ph.D. degree with the Digital Circuits and Systems group of Luca Benini. His research interests include transprecision computing as well as near- and in-memory processing.
\end{IEEEbiography}

\begin{IEEEbiography}[{\includegraphics[width=1in,height=1.25in,clip,keepaspectratio]{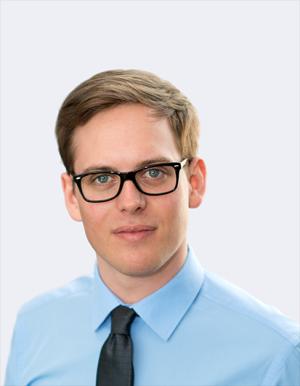}}]{Florian Zaruba}
  received his BSc degree from TU Wien in 2014 and his MSc from the Swiss Federal Institute of Technology Zurich in 2017. He is currently pursuing a Ph.D. degree at the Integrated Systems Laboratory. His research interests include the design of very large-scale integrated circuits and high-performance computer architectures.
\end{IEEEbiography}

\begin{IEEEbiography}[{\includegraphics[width=1in,height=1.25in,clip,keepaspectratio]{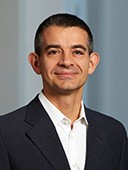}}]{Luca Benini}
    holds the chair of digital Circuits and systems at ETHZ and is Full Professor at the Universita di Bologna. Dr. Benini's research interests are in energy-efficient computing systems design, from embedded to high-performance. He has published more than 1000 peer-reviewed papers and five books. He is a Fellow of the ACM and a member of the Academia Europaea. He is the recipient of the 2016 IEEE CAS Mac Van Valkenburg Award and the 2020 EDAA Achievement Award.
\end{IEEEbiography}







\end{document}